\documentclass[11pt,a4paper]{article}
\pdfoutput=1 
\usepackage[dvipdfmx]{graphicx}
\usepackage[dvipdfmx]{color} % for output of png file

\usepackage{amsmath,amssymb,amsfonts,color,scalefnt}
\usepackage{epsfig,amsthm}
\usepackage{soul}
\usepackage[makeroom]{cancel}
\usepackage{mathrsfs}
\usepackage{epstopdf}
\usepackage[utf8]{inputenc}
\usepackage{hyperref}
\usepackage{slashed}
\usepackage{pict2e}
\usepackage{tikz}
\usepackage{cite}
\usepackage{float}
\usepackage[percent]{overpic}
\usepackage{multirow}
\usepackage{physics}
\usepackage{booktabs}
\usepackage{mwe}
\usepackage{subfig}
\usepackage [autostyle, english = american]{csquotes}
\usepackage{mathtools}
\MakeOuterQuote{"}

\newcommand{\al}{\alpha}
\newcommand{\beq}{\begin{eqnarray}}
\newcommand{\eeq}{\end{eqnarray}}
\newcommand{\ba}{\begin{eqnarray}}
\newcommand{\ea}{\end{eqnarray}}
\newcommand{\be}{\begin{equation}}
\newcommand{\ee}{\end{equation}}
\newcommand{\bpmatrix}{\begin{pmatrix}}
\newcommand{\epmatrix}{\end{pmatrix}}

\newcommand{\orderO}[1]{\mathcal{O}(#1)}

\newcommand{\OS}{\text{OS}}

\newcommand{\slashq}{\slashed{q}}

\newcommand{\doublet}[2]{\begin{pmatrix} #1 \\ #2 \end{pmatrix}}

\renewcommand{\Re}{\text{Re}}
\renewcommand{\Im}{\text{Im}}

\newcommand{\mueff}{\mu_{\text{eff}}}

\newcommand{\diag}{\text{diag}}
\newcommand{\bea}{\begin{eqnarray}}
\newcommand{\eea}{\end{eqnarray}}
\newcommand{\crn}{\nonumber \\}
\renewcommand{\braket}[1]{\left(#1\right)}

\newcommand{\calR}{{\mathcal R}}
\newcommand{\calM}{{\mathcal M}}
\newcommand{\fr}{\frac}

\newcommand{\figref}[1]{Fig.~\ref{#1}}
\renewcommand{\eqref}[1]{Eq.~(\ref{#1})}

\newcommand{\tab}[1]{Table~\ref{#1}}

\newcommand{\ssect}[1]{Subsection~\ref{#1}}

\newcommand{\appen}[1]{Appendix~\ref{#1}}

\newcommand{\ti}{\tilde}
\renewcommand{\abs}[1]{\left|#1\right|}
\newcommand{\calL}{{\cal L}}
\newcommand{\ISS}{\text{ISS}}

\newcommand{\tn}{{\tilde{\nu}}}
\newcommand{\tN}{{\tilde{N}}}
\newcommand{\tX}{{\tilde{X}}}
\newcommand{\Usn}{{U_{\tilde{\nu}}}}
\newcommand{\half}{\frac{1}{2}}
\newcommand{\vp}{\varphi}
\newcommand{\gev}{~\text{GeV}}

\usepackage[capitalise, english]{cleveref}
\crefname{section}{Sec.}{Secs.}
\crefname{table}{Tab.}{Tabs.}

\graphicspath{{figs/}}
\makeatletter
\def\input@path{{chapters/}}
\makeatother

\begin{document}
\vspace{1cm}

\title{
\vspace*{-3cm}
\phantom{h} \hfill\mbox{\small KA-TP-22-2022} 
\\[1cm]
\textbf{Leptonic Anomalous Magnetic and Electric Dipole Moments in the CP-violating NMSSM with and without Inverse Seesaw Mechanism}
}

\author{
Thi Nhung Dao$^{1}$\footnote{E-mail:
  \texttt{nhung.daothi@phenikaa-uni.edu.vn}}\;,
Duc Ninh Le$^{1}$\footnote{E-mail:
  \texttt{ninh.leduc@phenikaa-uni.edu.vn}}\;,
Margarete M\"uhlleitner$^{2}$\footnote{E-mail:
  \texttt{margarete.muehlleitner@kit.edu}}
\\[9mm]
{\small\it
$^1$Faculty of Fundamental Sciences, PHENIKAA University, Hanoi 12116,
Vietnam} \\
{\small\it
$^2$Institute for Theoretical Physics, Karlsruhe Institute of Technology,
76128 Karlsruhe, Germany} 
}
\maketitle

\begin{abstract}
The new results on the muon anomalous magnetic moment (AMM) published by
Fermilab in 2021, did not lead to a reduction of its long-pending deviation from
the Standard Model (SM) value by more than 4$\sigma$. The
explanation of this discrepancy by adding new particles to the theory
puts many new physics models under tension when combined with the null
results of the LHC direct searches for new particles. In this paper,
we investigate the CP-violating Next-to-Minimal Supersymmetric extension of the SM
(NMSSM) with and without an inverse seesaw mechanism. We compute the
one-loop supersymmetric contributions to the AMM
and the two-loop Barr-Zee-type diagrams with effective Higgs couplings
to photons for the leptonic electric dipole moments (EDMs). The effects
of the extended (s)neutrino sector on the muon AMM and on the mass of
the SM-like Higgs boson can be significant. Complex phases can have an important
impact on the AMM. On the other hand, the stringent limits from the EDMs
on the complex phases have to be taken into account. 
Our calculations have been implemented in the Fortran codes
{\tt NMSSMCALC} and {\tt NMSSMCALC-nuSS} which are publicly available. 
Besides the leptonic AMMs and EDMs, these programs can 
compute the Higgs boson masses and mixings, together with Higgs boson 
decay widths and branching ratios taking into account the most
up-to-date higher-order corrections in the NMSSM with and without
inverse seesaw mechanism. 
\end{abstract}

\thispagestyle{empty}
\vfill
\pagebreak

%%%%%%%%%%%%%%%%%%%%%%%%%%%%%%%%%%%%%%%%%%%%%%%%%%%%%%%%%%%%%%
\section{Introduction}
\label{sec:intro}
%%%%%%%%%%%%%%%%%%%%%%%%%%%%%%%%%%%%%%%%%%%%%%%%%%%%%%%%%%%%%%
At the beginning of 2021, the Fermilab Muon $g-2$ collaboration 
reported their first result \cite{Muong-2:2021ojo} of the muon 
anomalous magnetic moment (AMM) $a_\mu \equiv (g_\mu- 2)/2$,  
\be 
a_\mu^{\text{FNAL}} = (11659204.0\pm 5.4)\times 10^{-10} ,
\ee
which is consistent with the previous measurement by the E821
experiment at BNL with \cite{Muong-2:2006rrc}
 \be a_\mu^{\text{BNL}} = (116 592 08.9 \pm 6.3)\times 10^{-10}.\ee
The combined result, $a_\mu^{\text{exp}}= (11659206.1\pm 4.1)\times 10^{-10}$ compared with the theoretical prediction of the Standard Model (SM) \cite{Aoyama:2020ynm} 
 \be a_\mu^{\text{SM}} = (116 591 81.0 \pm 4.3)\times 10^{-10} \ee
leads to a deviation,
\be \Delta a_\mu\equiv a_\mu^{\text{exp}} -a_\mu^{\text{SM}}= (25.1 \pm 5.9)\times 10^{-10} ,\label{eq:anomaly}  \ee
at the $4.2\sigma$ level. The SM result consists of the pure
QED, electroweak and hadronic contributions. The
pure QED contribution has been evaluated up
to ${\cal O}(\alpha^5)$ \cite{Aoyama:2012wk} with negligible uncertainty,
the electroweak correction has been computed up to leading three-loop order with
less than one percent of uncertainty, see \cite{Aoyama:2020ynm} and references therein, and is suppressed by the ratio
$m_\mu^2/M_W^2$ where $m_\mu$ and $M_W$ are the mass of the muon and the
$W$ boson, respectively. The largest uncertainty comes from the
hadronic contributions which are calculated using non-pertubative
methods. Very recently, the hadronic light-by-light contribution was
computed by using lattice QCD 
\cite{Chao:2021tvp} and slightly reduced the significance of the
  anomaly. 

The anomaly is tantalizing in view of new physics at the weak scale
\cite{Czarnecki:2001pv}. Most of the models which  
try to explain this discrepancy tend to extend the electroweak sector to include
additional corrections, $a_\mu^{\text{new}}$.  In the Minimal
Supersymmetric extension of the SM (MSSM), besides the two Higgs doublets 
$H_u$ and $H_d$, there are additional fields given by the superpartners of the muons, Higgs 
bosons and gauge bosons that interact directly with the muons. They
enter the one-loop diagrams that contribute to  $a_\mu^{\text{new}}$. The new 
contributions depend on the ratio  $m_\mu^2/M_S^2 $, where $M_S$
represents the mass scale of the supersymmetric (SUSY) particles and
the muon Yukawa coupling $y_\mu =\sqrt{2} m_\mu/(v\cos\beta)$. Here
$v$ denotes the vacuum expectation value given in terms of the two
vacuum expectation values $v_u$ and $v_d$ of the two Higgs doublets
$H_u$ and $H_d$, respectively, $v= \sqrt{v_u^2+ v_d^2}$, and $ \tan\beta\equiv 
v_u/v_d$. The new contribution $a_\mu^{\text{new}}$ can be
significant when $M_S$ is small and/or $\tan\beta$ becomes large. The
non-observation of SUSY particles at the LHC, however, pushes the SUSY
mass scale $M_S$  to the TeV range. Moreover, the Higgs signals
measured at the LHC require the SM-like Higgs couplings to be close to
the ones of the SM, and therefore $\tan\beta$ should not be large. 
 Furthermore, the SM-like Higgs should be the 
  $h_u$-dominated Higgs boson so that it couples with
    a SM-like coupling to the top quarks.   
These requirements constrain the value of $a_\mu^{\text{new}}$. 
 
The Next-to-Minimal Supersymmetric SM (NMSSM)  contains an additional complex singlet superfield 
\cite{Fayet:1974pd,Barbieri:1982eh,Dine:1981rt,Nilles:1982dy,Frere:1983ag,Derendinger:1983bz,Ellis:1988er,Drees:1988fc,Ellwanger:1993xa,Ellwanger:1995ru,Ellwanger:1996gw,Elliott:1994ht,King:1995vk,Franke:1995tc,Maniatis:2009re,Ellwanger:2009dp}.
 Its scalar  component can mix with 
the scalar components of the two Higgs
doublet superfields which results in five neutral scalar Higgs boson
states. Although the LHC Higgs data has pushed the mass of 
the dominantly doublet-like scalar/pseudoscalar Higgs states,
$h_d/a_d$, into the TeV range it still allows for the singlet-like
Higgs boson masses to be in the GeV range. This makes the NMSSM an
interesting candidate for Higgs physics beyond the SM. As for the
muon AMM, one  expects a similar contribution
from the electroweakino sector as in the MSSM. A noticeable difference
may come from the contribution of a singlet-like Higgs boson with
  a mass of a few GeV. However, the one and two-loop 
light Higgs contributions are of opposite sign and therefore interfere destructively as
 shown in \cite{Krawczyk:2002df,Domingo:2008bb}. 
When the (s)neutrino sector of the NMSSM is
 extended to include six singlet leptonic superfields ($\hat N_i,\hat
 X_i$, $i=1,2,3$),  the three very  small neutrino masses 
 can then be generated through the inverse
seesaw mechanism \cite{Mohapatra:1986aw,PhysRevD.34.1642,Bernabeu:1987gr}. This extension of 
the NMSSM was first discussed in \cite{Gogoladze:2008wz}. The
$h_u$-like Higgs boson now has interactions with the left-handed
doublet neutrinos $\nu_L^i$ and the new singlet fermionic components
$N^i$, and also with  their scalar partners, that are proportional to
the neutrino Yukawa couplings. 
 These can  induce new one-loop contributions to
 the Higgs boson masses as shown 
 in \cite{Gogoladze:2012jp,Wang:2013jya,Dao:2021vqp}. This extension
 gives rise to the mixing  
 between left-handed doublet sneutrinos $\ti\nu_L^i$ with 
   the right-handed ones so that the
 sneutrino masses can be rather light.  This opens
 the possibility that the lightest sneutrinos can be a feasible Dark
 Matter candidate, as shown in \cite{Cao:2019aam,Cao_2020}. The
 extended sneutrino sector also gives rise to a new one-loop  
 contribution to the AMM of the charged leptons, as shown in \cite{Cao:2019evo,Cao:2021lmj}.   

In this study we compute and subsequently discuss
the full one-loop SUSY contributions to the leptonic AMM 
and electric dipole moment (EDM) in the NMSSM and a variant of the NMSSM  
 with inverse seesaw mechanism (abbreviated as NMSSM-nuSS)
 taking into account non-vanishing CP-violating phases. We further
 include contributions from the two-loop Barr-Zee-type diagrams with
 effective $h\gamma\gamma$ couplings. We show in this study the
 correlation between the impacts of the extended (s)neutrino sector on
 the muon AMM and on the loop-corrected
 $h_u$-like Higgs boson mass. The impacts can be significant
 simultaneously. In the regions where a positive SUSY contribution to the muon 
  AMM is necessary to explain the anomaly, the
  one-loop contributions from the extended (s)neutrino sector to the $h_u$-like Higgs
  boson mass can become
  negative since the sneutrino contributions dominate over the neutrino
  contributions. We also study the effects of the complex phases on  
 the muon AMM in both models. All these
 contributions to the AMM and to the EDM of the charged leptons have been implemented in our two published 
 Fortran codes {\tt NMSSMCALC}
 \cite{Baglio:2013iia,Graf:2012hh,Muhlleitner:2014vsa,Dao:2019qaz,Dao:2021khm}
 and {\tt NMSSMCALC-nuSS} \cite{Dao:2021vqp} which compute the Higgs
 boson masses and mixings, together with Higgs boson decay widths and
 branching ratios taking into account the most up-to-date higher-order
 corrections. The codes can be downloaded from the url:
\begin{center}
\nolinkurl{https://www.itp.kit.edu/~maggie/NMSSMCALC/}
\end{center}
and
\begin{center}
\nolinkurl{https://www.itp.kit.edu/~maggie/NMSSMCALC-nuSS/}
\end{center}

The paper is organised as follows. 
Section \ref{sec:model} introduces the models and our notations. In
Section \ref{sec:calculationframework} we present our computation and
analytical expressions of the 
one-loop and two-loop contributions to the leptonic AMM and EDM. 
The set-up of the calculation and the numerical analysis are given in 
Sec.~\ref{sec:results}. We conclude in section \ref{sec:conclusions}. 

%%%%%%%%%%%%%%%%%%%%%%%%%%%%%%%%%%%%%%%%%%%%%%%%%%%%%%%%%%%%%%
\section{The Complex NMSSM and the NMSSM with Inverse Seesaw Mechanism  }
\label{sec:model}
%%%%%%%%%%%%%%%%%%%%%%%%%%%%%%%%%%%%%%%%%%%%%%%%%%%%%%%%%%%%%%
The difference between the complex NMSSM and the complex  NMSSM with
inverse seesaw mechanism
manifests itself mainly in the neutrino and sneutrino sectors. We start with
a short description of the complex NMSSM to introduce the model
parameters. We follow the same notation which has been used in  our
previous studies
\cite{Graf:2012hh,Baglio:2013iia,Muhlleitner:2014vsa,Dao:2019qaz,Dao:2021khm}. 
 The complex NMSSM superpotential is given by ($i,j=1,2$)
\begin{align}
    \mathcal{W}_{\text{NMSSM}} = 
	\epsilon_{ij} [y_e \hat{H}^i_d \hat{L}^j
\hat{E}^c + y_d \hat{H}_d^i \hat{Q}^j \hat{D}^c - y_u \hat{H}^i_u
\hat{Q}^j \hat{U}^c] -\epsilon_{ij} \lambda \hat S \hat H_d^i\hat H_u^j+\frac{1}{3}\kappa {\hat S}^3 \;,
\end{align}
with the quark and leptonic superfields $\hat{Q}$, $\hat{U}$, $\hat{D}$, $\hat{L}$, $\hat{E}$,
and the Higgs doublet
superfields $\hat{H}_d$, $\hat{H}_u$ and the singlet superfield
$\hat{S}$ and the totally antisymmetric tensor $\epsilon_{12}= \epsilon^{12}=1$. Charge conjugated fields are denoted by the superscript $c$. Color and generation indices have been suppressed for the sake of clarity. The Yukawa couplings $y_u,\
y_d$ and $y_e$ are taken as diagonal 3$\times$3 matrices in the flavour space. 
The coupling parameters $\lambda$ and $\kappa$ are complex numbers
in the CP-violating NMSSM.
The soft SUSY-breaking Lagrangian reads
\begin{align}\label{eq:breaking_term}\notag
\mathcal{L}_{\rm soft,NMSSM} = & -m_{H_d}^2 H_d^\dagger H_d - m_{H_u}^2
H_u^\dagger H_u -
m_{\tilde{Q}}^2 \tilde{Q}^\dagger \tilde{Q} - m_{\tilde{L}}^2 \tilde{L}^\dagger \tilde{L}
- m_{\tilde{u}_R}^2 \tilde{u}_R^*
\tilde{u}_R - m_{\tilde{d}_R}^2 \tilde{d}_R^* \tilde{d}_R
\nonumber      \\\nonumber
& - m_{\tilde{e}_R}^2 \tilde{e}_R^* \tilde{e}_R - (\epsilon_{ij} [y_e A_e H_d^i
\tilde{L}^j \tilde{e}_R^* + y_d
A_d H_d^i \tilde{Q}^j \tilde{d}_R^* - y_u A_u H_u^i \tilde{Q}^j
\tilde{u}_R^*] + \mathrm{h.c.})      \\
& -\frac{1}{2}(M_1 \tilde{B}\tilde{B} + M_2
\tilde{W}_j\tilde{W}_j + M_3 \tilde{G}\tilde{G} + \mathrm{h.c.}) \\ \nonumber
& - m_S^2 |S|^2 +
(\epsilon_{ij} \lambda
A_\lambda S H_d^i H_u^j - \frac{1}{3} \kappa
A_\kappa S^3 + \mathrm{h.c.}) \;.
\end{align}
 The $H_{u,d}$ are two scalar Higgs doublets fields, $S$ a scalar
 singlet field, $\tilde{Q}$ scalar squark doublets, $\tilde{L}$ scalar slepton doublets, 
$\tilde{u}_R$ and $\tilde{d}_R$ scalar squark singlet fields, and
$\tilde{e}_R$ a scalar slepton singlet field. The soft
SUSY-breaking gaugino mass parameters $M_k$ ($k=1,2,3$) of the bino,
wino and gluino fields $\tilde{B}$, $\tilde{W}_l$ ($l=1,2,3$) and
$\tilde{G}$ as well as the soft SUSY-breaking trilinear couplings
$A_x$ ($x=\lambda,\kappa,u,d,e$) are complex in the CP-violating
NMSSM.

After electroweak symmetry breaking, the Higgs boson fields can be
expanded around their vacuum expectation values (VEVs) $v_u$, $v_d$,
and $v_s$, respectively,  
\begin{equation}
    H_d = \doublet{\frac{v_d + h_d +i a_d}{\sqrt 2}}{h_d^-}, \,\, 
    H_u = e^{i\varphi_u}\doublet{h_u^+}{\frac{v_u + h_u +i a_u}{\sqrt 2}},\,\,
    S= \frac{e^{i\varphi_s}}{\sqrt 2} (v_s + h_s + ia_s)\, ,
   \label{eq:vevs}
\end{equation}
with the CP-violating phases $\varphi_{u,s}$ and  we
obtain the tree-level spectrum of the Higgs sector. The relation
  to the SM VEV $v\approx 246.22$~GeV is given by
\begin{equation}
   v^2 = v_u^2 +v_d^2  
\end{equation} 
and we define the mixing angle $\tan\beta$ as
\begin{equation}
   \tan\beta = \fr{v_u}{v_d} \;.
   \label{eq:tan_beta}
\end{equation}
The effective $\mu$ parameter is given by
\be 
\mu_{\text{eff}} = \fr{\lambda v_s e^{i\varphi_s}}{\sqrt 2}.
\ee 
Besides the gauge bosons, quarks, charged leptons, and three left-handed
neutrino fields as in the SM, we have an extended Higgs spectrum and
new SUSY particles, in particular: 
\begin{itemize}
\item 
The CP-even and CP-odd Higgs interaction states $(h_{d,u,s},
a_{u,d,s})$ mix to form five CP indefinite Higgs mass eigenstates $h_{i}$ 
($i=1,...,5$), with their masses per convention ordered as   
$m_{h_1}<m_{h_2}<m_{h_3}<m_{h_4}<m_{h_5}$, and one neutral
Goldstone boson $G^0$. We use a two-fold rotation 
 to rotate from the interaction to the mass eigenstates, 
\beq 
 (h_d, h_u,h_s, a, a_s,G^0)^T&=& \mathcal{R}^G(\beta)\, (h_d, h_u,
h_s, a_d, a_u, a_s)^T,\\
 (h_1,h_2,h_3,h_4,h_5,G^0)^T& = & \mathcal{R}^H\, (h_d, h_u,
h_s, a, a_s,G^0)^T,
\eeq
 where the first rotation matrix $\mathcal{R}^G$ with one rotation
 angle $\beta$ singles out the neutral Goldstone boson and the second
 rotation matrix $\mathcal{R}^H$ rotates the five interaction states
 $(h_d, h_u, h_s, a, a_s)$ to the five mass eigenstates
 $(h_1,h_2,h_3,h_4,h_5)$.
\item The charged Higgs interaction states $h_d^\pm,h_u^\pm$
  constitute the charged Higgs bosons $H^\pm$ with mass $M_{H^\pm}$
  and the charged Goldstone bosons $G^\pm$. 
\item The fermionic partners of the neutral Higgs bosons, the neutral
higgsinos $\tilde{H}_u$, $\tilde{H}_d$ and the singlino
$\tilde{S}$, mix with the neutral gauginos $\tilde{B}$ and
$\tilde{W}^3$, resulting in five neutralinos denoted as
$\tilde{\chi}^0_i$, $(i=1,...,5)$. 
 The mass ordering of the  $\tilde{\chi}^0_i$ is chosen as 
$m_{\tilde{\chi}^0_1}\leq...\leq m_{\tilde{\chi}^0_5}$ and 
 the rotation matrix 
$N$ transforms the fields $(\tilde{B},\ \tilde{W}^3,\ \tilde{H}_d,\ \tilde{H}_u,
\tilde{S})^T$  into the mass eigenstates. 
\item The two chargino mass eigenstates,
\beq
\tilde{\chi}_i^+ = \left( \begin{array}{c} \ti{\chi}_{L_i}^+ \\[0.1cm]
                             \overline{\ti
                            \chi_{R_i}^-} \end{array}\right) \;, \quad
                      i=1,2 \;,
\eeq
are obtained from the rotation of the interaction states, given by
  the charged Higgsinos $\tilde{H}^\pm_d$, $\tilde{H}^\pm_u$ and the
  charged gauginos $\tilde{W}^\pm$, to the mass eigenstates. This is
  done by a bi-unitary transformation with the two $2\times 2$ unitary
  matrices $V^\chi$ and $U^\chi$,
 \begin{align}
 \ti{\chi}_L^+=V^{\chi}(\tilde{W}^+, \tilde{H}^+_u)^T,\ \ \
  \ti{\chi}_R^-=U^{\chi}(\tilde{W}^-, \tilde{H}^-_d)^T. 
 \end{align}
 \item The scalar partners of the left- and right-handed up-type
   quarks are denoted by $\ti u_{L/R}^i$, of the down-type quarks by $\ti
   d^i_{L/R}$, and of the charged leptons by $\ti l^i_{L/R}$
   ($i=1,2,3$). We do not include flavor mixing. Within each flavour
   the left- and right-handed scalar fermions with same electric
     charge mix and they are rotated to the mass eigenstates by a
   unitary matrix $U^{\ti f}$. 
   \item There are three scalar partners of the left-handed neutrinos,
     denoted as $\ti \nu_i$ ($i=1,2,3$) with their masses given by
\be 
m^2_{\ti \nu_i} = \fr{1}2 M_Z^2 c_{2\beta} + m_{\ti{L}_i}^2 \;
, 
\ee
 where the short hand notation $c_{x} \equiv \cos(x), s_{x} \equiv
 \sin(x), t_{x} \equiv \tan(x)$ is used in this paper and the second
 term comes from the soft SUSY-breaking Lagrangian in \eqref{eq:breaking_term}.
\end{itemize} 

The complex NMSSM with inverse seesaw mechanism is obtained from the
complex NMSSM by including six gauge-singlet  chiral superfields
$\hat{N}_i$, $\hat X_i$ ($i=1,2,3$) that carry lepton number. We
follow the same notation as in our previous investigation of the loop
corrections to the neutral Higgs boson masses presented in
\cite{Dao:2021vqp}. The superpotential of the model reads ($i,j=1,2$)
\begin{align} 
\mathcal{W}_{\text{NMSSM-nuSS}} =\mathcal{W}_{\text{NMSSM}}- y_\nu \epsilon_{ij}  \hat{H}_u^i \hat{L}^j \hat{N}^c +  \lambda_X  \hat S \hat X
\hat X + \mu_X  \hat X  \hat{N}^c \,, \label{eq:wnmssm}
\end{align}
where the neutrino Yukawa coupling $y_\nu$ and the coupling
$\lambda_X$ are  $3\times 3$ complex matrices in general, and the
superscript $c$ denotes the charge conjugation. 
The $3\times 3$ matrix $\mu_X$ is the only parameter
with the dimension of mass in the superpotential so that it can be of the
order of the SUSY-conserving mass scale and is naturally large. 
The soft SUSY-breaking NMSSM Lagrangian respecting the gauge symmetries
and the global $\mathbb{Z}_3$ symmetry reads (the assignment of 
the $\mathbb{Z}_3$ charges is provided in \cite{Dao:2021vqp}) 
\begin{align}
\mathcal L_{\text{NMSSM-nuSS}}^{\text{soft}} =& \mathcal L_{\text{NMSSM}}^{\text{soft}}
+ (\epsilon_{ab} y_\nu  A_{\nu}  H_u^a \ti{L}^b\ti N^*  
%\crn
%& 
+ \lambda_X A_X  S\ti X \ti X +\mu_X B_{\mu_X}  \ti X \ti N^* + h.c.)
   \crn
&  - \ti m^2_X \abs{\ti X}^2 - \ti m^2_N \abs{\ti N}^2   \; ,
\end{align}
which introduces
 the soft SUSY-breaking trilinear couplings  $ A_\nu, A_X$, the soft
 SUSY-breaking masses $\ti m_X^2,\ti m_N^2$, and the soft SUSY-breaking bilinear mass
$B_{\mu_X}$.

In the neutral leptonic sector, the three left-handed neutrinos
$\nu_{L_i}$ mix with the six leptonic component fields of the six singlet 
superfields $ \hat{N}_i^c, \hat X_i$, $i=1,2,3$, and the mass term in
the Lagrangian reads 
\be 
\calL_{\text{mass}}^\nu =-\fr{1}{2} \bpmatrix \nu_L & N^c & X
\epmatrix  M_\ISS^\nu  \bpmatrix \nu_L\\ N^c\\ X \epmatrix ,
\ee
where the mixing mass matrix is given by
\be M_{\ISS}^\nu = \bpmatrix  0 & M_D & 0 \\
M_D^T & 0 & \mu_X \\
0  & \mu_X^T & M_X \epmatrix , \label{eq:neumassmatrix}\ee 
where blocks $M_D, \mu_X$ and $M_X$ are $3\times 3$ matrices with $
\mu_X $ defined in Eq.~(\ref{eq:wnmssm}) and 
\be 
M_D =\fr{v_u e^{i\varphi_u}}{\sqrt{2}} y_\nu, \quad   M_X =
\fr{v_se^{i\varphi_s}}{\sqrt{2}} (\lambda_X + \lambda_X^T) \;. \label{eq:lambdaxinmasses}
\ee
Diagonalizing the neutrino mass matrix with a unitary rotation matrix
$U^\nu$, one obtains nine neutrino mass eigenstates with their masses
$m_{\nu_i}$ $(i=1,...,9)$ being sorted in ascending order.  
By exploiting the fact that all matrix elements 
of $M_D$ and $M_X$ are much smaller than the eigenvalues of $\mu_X$,
the $3\times 3$ light neutrino mass matrix can be expressed at leading order as 
\be 
M_{\text{light}} = M_D M_N^{-1} M_D^T\,, \; \mbox{ with } \; M_N= \mu_X M_X^{-1}
\mu_X^T\,,\label{eq:mlight}
\ee
and then can be diagonalized by  the Pontecorvo-Maki-Nakagawa-Sakata (PMNS) matrix $U_{\text{PMNS}}$,
 \bea
  U_{\text{PMNS}}^* M_{\text{light}} U_{\text{PMNS}}^\dagger = m_\nu \;, \label{eq:LOexpansion} \quad 
  m_\nu= \diag({m_{\nu_1}},{m_{\nu_2}},{m_{\nu_3}}).
\eea  
In order to reproduce the light neutrino oscillation data, two different parameterizations
 have been considered. In the so-called Casas-Ibarra parameterization
 \cite{Casas:2001sr}, $M_D$ is computed from the 
 relation
\bea 
M_D =  U_{\text{PMNS}}^T \sqrt{m_\nu}  R  \sqrt{{\calM}_N} V_\nu, \label{eq:MDdefinition} \quad 
{\calM}_N=\diag({M_{N_1}},{M_{N_2}},{M_{N_3}})=  V^*_\nu M_N
V^\dagger_\nu \;,
\eea
with $R$ being a complex orthogonal matrix and
$V_\nu$ a unitary matrix diagonalizing $M_N$. The $y_\nu$ are then
obtained from \eqref{eq:lambdaxinmasses}. The other possibility is to use the
$\mu_X$-parameterization \cite{Arganda:2014dta} in which $M_X$ is computed from the relation
\bea M_X= \mu_X^T M_D^{-1} U_{\text{PMNS}}^* m_\nu  U_{\text{PMNS}}^\dagger  M_D^{T,-1} \mu_X, \eea
where $M_D$ is calculated from the input $y_\nu$. 

In the sneutrino sector, each sneutrino field is split up into its
CP-even and CP-odd components as  
\begin{align}
\tn &= \frac{1}{\sqrt{2}} \braket{\tn_+ + i \tn_-},\\
\tN^* &= \frac{1}{\sqrt{2}} \braket{\tN_+ + i \tN_-},\\
\tX &= \frac{1}{\sqrt{2}} \braket{\tX_+ + i \tX_-}.
\end{align}
The mass term in the basis $ \psi = (\tn_+,\tN_+,\tX_+,\tn_-,\tN_-,\tX_-)^T $ (generation indices are suppressed) is given by
\be 
{\cal L}= \fr 12 \psi^T M_{\tn}\psi\;,
\ee
where the mass matrix $ M_\tn $ is an $18\times 18$ symmetric matrix
that can be found in \appen{appen:sneumass}.  An orthogonal matrix $
\Usn $ can be used to obtain the masses of the sneutrinos as 
\be
\diag \braket{m^2_{\tilde{n}_1},\cdots,m^2_{\tilde{n}_{18}}} = \Usn M_\tn \Usn^T,
\ee
where their mass values are ordered as $ m^2_{\tilde{n}_1} \leq \cdots \leq m^2_{\tilde{n}_{18}} $.

%%%%%%%%%%%%%%%%%%%%%%%%%%%%%%%%%%%%%%%%%%%%%%%%%%%%%%%%%%%%%%
\section{SUSY Contributions to the Leptonic AMM and EDM}
\label{sec:calculationframework}
%%%%%%%%%%%%%%%%%%%%%%%%%%%%%%%%%%%%%%%%%%%%%%%%%%%%%%%%%%%%%%
The SUSY contributions to the leptonic AMM $a_l$ and EDM $d_l$ ($l=e,\mu,\tau$ ) 
can be calculated in perturbation theory by considering the matrix
element decomposed into a relativistic covariant form,
\bea
\langle l(p_2) | j^\mu(q)|l(p_1)\rangle&=& -ie \bar u(p_2)\bigg[
\left(\gamma^\mu - \fr{\slashq q^\mu}{q^2}\right) (\bar{F}_L(q^2)P_L + \bar{F}_L^{*}(q^2)P_R) \label{eq:formfactor}\\
&&+ \fr{i \sigma^{\mu\nu}q_\nu}{2m_l}(F_L(q^2)P_L + F_L^{*}(q^2)P_R) 
 + \fr{q_\mu}{m_l}(\bar{\bar{F}}_L (q^2)P_L + \bar{\bar{F}}_L^{*}
 (q^2)P_R)  \bigg] u(p_1)\nonumber
\eea
where $\sigma^{\mu\nu}=\fr i2[\gamma^\mu,\gamma^\nu]$, $P_{L/R}= \fr{1\mp
  \gamma_5}{2}$, $q=p_2-p_1$, $m_l$ is the lepton mass  and $u(p)$
denotes the Dirac spinor. The form factors $F_L$, $\bar{F}_L$,
$\bar{\bar{F}}_L$ are functions of $q^2$ and other parameters of the
model.  
The operator $ \sigma^{\mu\nu}q_\nu$ is called dipole matrix operator. 
In the static limit ($q^\mu \to 0$) we have \cite{Jegerlehner:2009ry}:
\be a_l= \Re[F_L(0)], \quad d_l=\fr{e}{2m_l}\Im[F_L(0)].\ee
In our computation we will use this generic form for both the AMM and 
EDM keeping all possible complex phases.

%%%%%%%%%%%%%%%%%%%%%%%%%%%%%%%%%%%%%%%%%%%%%%%%%%%%%%%%%%%%%
\subsection{One-Loop Contributions \label{sec:one-loopAMM}} 
%%%%%%%%%%%%
\begin{figure}[h]
    \centering
        \includegraphics[width=0.8\textwidth]{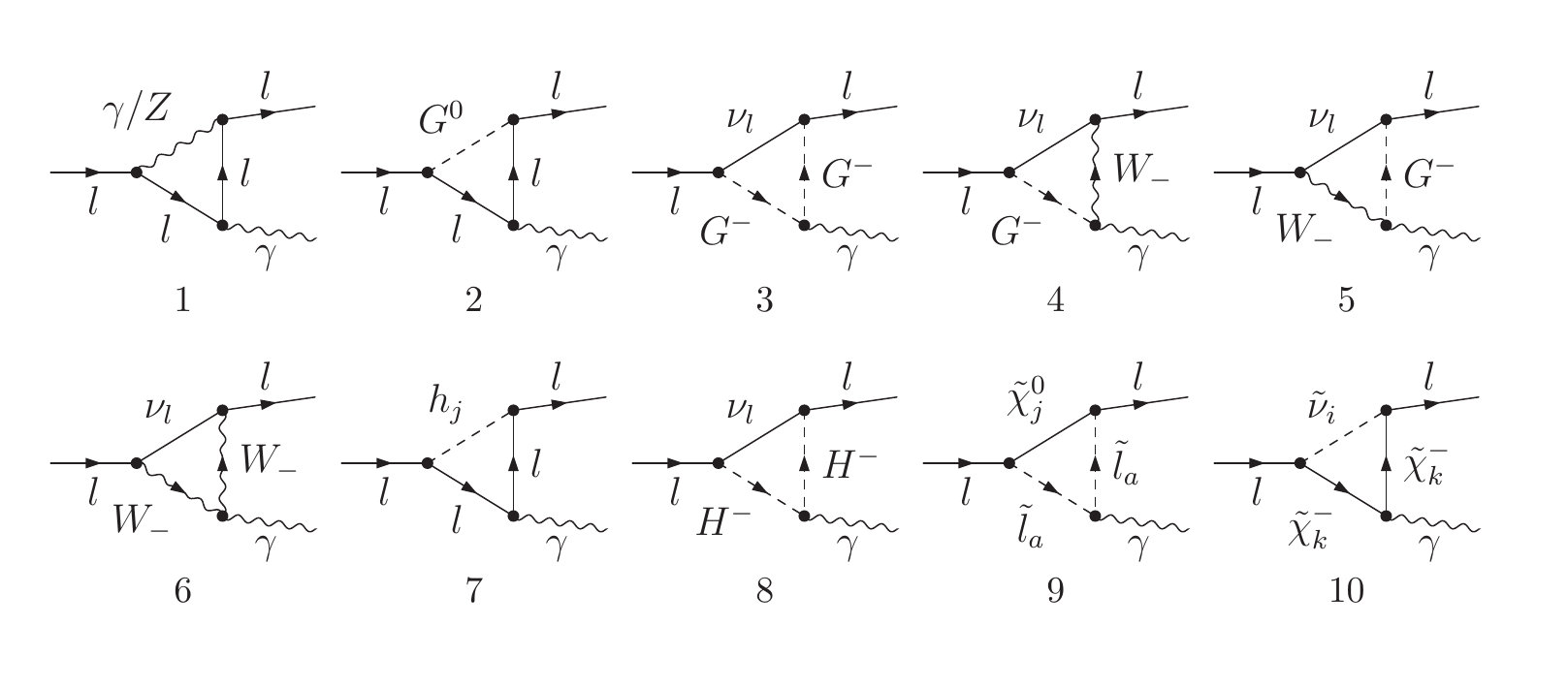}
    \caption{Generic one-loop Feynman diagrams contributing to
      the dipole matrix form factor.  
  } 
    \label{fig:OL-DPM}
\end{figure}
It is well known that the contributions to the dipole matrix
$\sigma^{\mu\nu}q_\nu$ require a chirality flip. 
Therefore contributions to $F_L(0)$ are proportional either to the mass of the external 
lepton or to the masses of the fermions running in the loop diagrams. 
In \figref{fig:OL-DPM}, we present all 
one-loop diagrams which contribute to $F_L(0)$ in the NMSSM-nuSS. 
Diagrams with neutral or charged Goldstone bosons occur explicitly 
because we work in the Feynman-'t Hooft gauge.  
The diagrams 1 and 2 are the same as in the SM. The diagram with the
photon exchange belongs to the QED contribution and the
diagrams with $Z$ and $G^0$ belong to the weak contribution. We
calculated these diagrams and recovered the results quoted in the
literature, see for example \cite{Jegerlehner:2009ry} and  
references therein. Since we donot account them in $a_l^{\text{new}}$
we do not present their explicit expressions here and will not mention them any more. 
The diagrams 3-6 belong to the $W$ contribution. In the SM, neutrinos are purely left-handed
and massless while in the NMSSM-nuSS we have three light active neutrinos and six sterile neutrinos.
 We denote the difference between the NMSSM-nuSS and the SM contributions with respect to the $W$ diagrams as
 \be F_L^{\text{new},W}= F_L^{\text{NM},W}-F_L^{\text{SM},W} \ee
 where ($l=e,\mu,\tau$)
 \bea  
F_L^{\text{NM},W} &=& -\fr{ m^2_l}{4 \pi^2v^2}\sum_{i=1}^{9} 
 \bigg[\fr 12 v^2 s_\beta^2 
 \bigg|\sum_{k=1}^3(U_{i(k+3)}^{\nu\;*} y^\nu_{lk})\bigg|^2 \left(
  C_2^{W,\nu_i} +  C_{12}^{W,\nu_i} +  C_{22}^{W,\nu_i} \right)\crn
&&
+\fr1{\sqrt2} s_\beta v m_{\nu_i}U_{il}^{\nu\;*} e^{i\varphi_u} \sum_{k=1}^3(U_{i(k+3)}^{\nu\;*} y^\nu_{lk})\left( C_0^{W,\nu_i} +C_1^{W,\nu_i} +C_2^{W,\nu_i}   \right)
 \\
&& +M_W^2 |U_{il}^{\nu}|^2 \left( -2 C_1^{W,\nu_i} + 2 C_{12}^{W,\nu_i} + 2
  C_{22}^{W,\nu_i}\right) 
  + m_l^2 |U_{il}^{\nu}|^2\left( C_1^{W,\nu_i} +  C_{12}^{W,\nu_i}+C_{11}^{W,\nu_i} \right)
  \bigg] \nonumber \eea
  and
  \bea  F_L^{\text{SM},W} =& -\fr{ m_l^2}{4 \pi^2v^2}\bigg[M_W^2 \left( -2  C_1^{W,\nu} + 2 \ti C_{12}^{W,\nu} + 2
 C_{22}^{W,\nu}\right) + m_l^2 \left(  C_2^{W,\nu} +  C_{12}^{W,\nu}+ C_{22}^{W,\nu} \right)
  \bigg] .\eea 
Note that we have introduced the abbreviations  $C_{i...}^{x,y}$ for the one-loop three-point integral coefficients which will be defined at the end of this section.
 Expanding $F_L^{\text{SM},W} $ with respect to $m_l/M_W$ we obtained
the first term in the expansion in accordance with the well-known
result in the literature, see for example \cite{Jegerlehner:2009ry}
and references therein,
\be 
a_l^{\text{SM},W}  = \fr{\sqrt 2 G_\mu m_l^2}{16 \pi^2}\fr{10}{3} \;,
\ee
where $G_\mu$ is the Fermi constant of the muon.
While $F_L^{\text{SM},W} $ is real in the SM, $F_L^{\text{NM},W}$ can be complex.
In the NMSSM without inverse seesaw mechanism, the neutrino sector 
is identical to the one of the SM, therefore
$F_L^{\text{new},W}$ vanishes.
The neutral Higgs boson contribution arises from diagram 7. In our
model there are five neutral Higgs bosons while in the SM there is only
one neutral state. We denote the new contribution from the
  neutral Higgs bosons as 
 \be 
F_L^{\text{new},H}= F_L^{\text{NM},H}-F_L^{\text{SM},H} \;,
\ee
 where 
 \bea  F_L^{\text{NM},H} &=& \fr{m_l^4}{8\pi^2v^2 } \sum_{j=1}^5\bigg[
  \fr{(\calR^H_{j1})^2}{c_\beta^2} \left( 2  C_1^{h_j,l}+2  C_2^{h_j,l} +
   C_{11}^{h_j,l}+ 2C_{12}^{h_j,l}+ C_{22}^{h_j,l}\right) \crn
   &&t_\beta^2(\calR^H_{j4} )^2 \left(  C_{11}^{h_j,l}+2C_{12}^{h_j,l}+C_{22}^{h_j,l}\right)  +2 i  t_\beta\fr{\calR^H_{j1}\calR^H_{j4}}{c_\beta}  \left(  
   C_{1}^{h_j,l} +C_{2}^{h_j,l}  \right)\bigg] \label{eq:NMHiggs} \eea
and
 \bea F_L^{\text{SM},H} =\fr{m_l^4}{8\pi^2v^2}\bigg[2  C_1^{H_{\text{SM}},l}+2  C_2^{H_{\text{SM}},l} +
   C_{11}^{H_{\text{SM}},l}+ 2C_{12}^{H_{\text{SM}},l}+ C_{22}^{H_{\text{SM}},l}\bigg] . \label{eq:SMHiggs}\eea
From \eqref{eq:NMHiggs} and \eqref{eq:SMHiggs} it is clear that the
Higgs contribution is suppressed by a factor of $m_l^2/M_W^2$ compared
to the $W$ and $Z$ contributions. The last three diagrams 8, 9, and 10
do not appear in the SM. They give rise to the charged Higgs,
neutralino and chargino contributions. Our calculations lead to the
following results 
\bea
 F_L^{\text{NM},H^\pm} &=&-\fr{ m_l^2}{16\pi^2 }\sum_{i=1}^{9} \bigg[2c_\beta^2 \bigg|\sum_{k=1}^3(U_{i(k+3)}^{\nu\;*} y^\nu_{lk})\bigg|^2(C_2^{H^\pm,\nu_i}+
 C_{12}^{H^\pm,\nu_i}+C_{22}^{H^\pm,\nu_i}  ) \label{eq:chargedHiggsCT}\\
 &&-\fr{2^{3/2}e^{i\varphi_u}  m_{\nu_i} s_\beta U_{il}^{\nu\;*}}{v} \sum_{k=1}^3(U_{i(k+3)}^{\nu\;*} y^\nu_{lk})
(C_{0}^{H^\pm,\nu_i} + C_{1}^{H^\pm,\nu_i} +C_{2}^{H^\pm,\nu_i}) \crn
&&
+ \fr{4m_l^2}{v^2}\abs{U_{il}^{\nu}}^2t_\beta^2( C_1^{H^\pm,\nu_i} +  C_{12}^{H^\pm,\nu_i} +  C_{11}^{H^\pm,\nu_i})  \bigg],\nonumber
\eea 
%%%%
\bea
F_{L}^{\text{NM},\tilde{\chi}^\pm}&=&\fr{m_l}{8 \pi^2} \sum_{i=1}^{18}\sum_{j=1}^{2} 
\bigg[
  m_l \left(\abs{g^{L}_{l\ti\chi_j^- \ti\nu_i} }^2 + \abs{g^{R}_{l
  \ti\chi_j^- \ti\nu_i} }^2\right)
   \left(C_1^{\ti \chi^\pm_j,\ti \nu_i}+C_{12}^{\ti \chi^\pm_j,\ti \nu_i}+
   C_{11}^{\ti \chi^\pm_j,\ti \nu_i}\right) \crn
  && +m_{\ti \chi^\pm_j} g^{R*}_{l\ti\chi_j^- \ti\nu_i} g^L_{l\ti\chi_j^- \ti\nu_i}  \left(C_1^{\ti \chi^\pm_j,\ti \nu_i}+C_2^{\ti \chi^\pm_j,\ti \nu_i}\right)  \bigg], \label{eq:olchargino}\eea 
 %%%%
\bea
F_{L}^{\text{NM},\tilde{\chi}^0}&=&\fr{m_l}{8 \pi^2} \sum_{a=1}^{2}\sum_{k=1}^{5} 
\bigg[
 - m_l \left(\abs{g^{L}_{l\ti\chi_k^0 \ti l_a} }^2 + \abs{g^{R}_{l\ti\chi_k^0 \ti l_a}
   }^2\right)
   \left(C_1^{\ti \chi_k^0,\ti l_a}+C_{12}^{\ti \chi_k^0,\ti l_a}+
   C_{11}^{\ti \chi_k^0,\ti l_a}\right) \crn
  && +m_{\ti \chi_k^0} g^{R*}_{l\ti\chi_k^0 \ti l_a} g^L_{l\ti\chi_k^0 \ti l_a }  \left(C_0^{\ti \chi_k^0,\ti l_a }+C_1^{\ti \chi_k^0,\ti l_a }+C_2^{\ti \chi_k^0,\ti l_a}\right)  \bigg]. \label{eq:olneutralino}\eea 
The left- and right-handed couplings of the charginos and the
neutralinos are defined in the interaction Lagrangian, 
\be 
\bar{l} ( i g^L_{l\ti\chi_j^- \ti\nu_i} P_L + i
g^R_{l\ti\chi_j^- \ti\nu_i} P_R ) \chi_j^- \ti\nu_i +
\bar{\chi}_k^0 ( i g^L_{l\ti\chi_k^0 \ti l_a} P_L + i
g^R_{l\ti\chi_k^0 \ti l_a } P_R ) l \ti l_a^* ,
\ee
where 
\begin{align}
	g^L_{l\ti\chi_j^- \ti\nu_i} &= \fr{m_l}{
                                          v c_\beta} \braket{
                                    U^{\ti\nu}_{il}
                            + i  U^{\ti\nu}_{i(l+9)}} U^{\chi\,*}_{j2},\\
	g^R_{l\ti\chi_j^- \ti\nu_i} &=
        -\fr{g_2V_{j1}^\chi}{\sqrt{2}}\braket{U^{\ti\nu}_{il}
                                          + i U^{\ti\nu}_{i(l+9) }
                                          }+\fr{1}{\sqrt 2}V_{j2}^\chi 
		 \braket{U^{\ti\nu}_{i(l+3)} + iU^{\ti\nu}_{i(l+12)}}
                                        y^{\nu*}_{ll} ,
\label{eq:charginoCL} \\
		g^L_{l\ti\chi_k^0 \ti l_a}&=\left( \fr{g_1}{\sqrt{2}} N_{k1}^*+\fr{g_2}{\sqrt{2}} N_{k2}^*\right)
		 U^{\ti l}_{a1 }
		+  \fr{\sqrt{2}m_l}{vc_\beta} N_{k3}^*  U^{\ti l}_{a2 }  , \\
	g^R_{l\ti\chi_k^0 \ti l_a}&= -\sqrt{2} g_1 N_{k1} U^{\ti l}_{a2 } - \fr{\sqrt{2}m_l}{v c_\beta}N_{k3} U^{\ti l}_{a1 },
\end{align}
where $g_1$ and $g_2$ are the gauge couplings of the $U(1)_Y$ and
$SU(2)_L$ gauge groups, respectively.
Note that we allow for lepton flavour mixing in the neutrino sector, therefore
 $y^\nu$ is a $3\times 3 $ complex matrix and $y^\nu_{ll}$ is the lth diagonal element with $l=1,2,3$ for electron, muon and tauon, respectively. 
 For the complex NMSSM without inverse seesaw mechanism,  one sets $y^\nu$ to zero and $U^\nu, U^{\ti\nu}$ to a $3\times 3$ unity matrix
in  the charged Higgs and chargino  contributions. The neutralino contribution is the same in both models.
In summary the new one-loop contributions from the weak sector of the NMSSM-nuSS
 to the leptonic AMM and EDM are given by
\bea  
F_L^{\text{new},1l} &=& F_L^{\text{new},W} + F_L^{\text{new},H} +F_{L}^{\text{NM},H^\pm} +F_{L}^{\text{NM},\tilde{\chi}^\pm}+F_{L}^{\text{NM},\tilde{\chi}^0},\\
a_l^{1l}&=& \Re[F_L^{\text{new},1l} ], \quad d_l^{1l}=\fr{e}{2m_l}\Im[F_L^{\text{new},1l} ].
\eea  
Finally, we define the  abbreviations for the
one-loop three-point integrals which have been used earlier in this
section,  
\begin{align}
    C_{i...}^{x,y} &= C_{i...} \braket{m_l^2,0,m_l^2, m_x^2,m_y^2,m_y^2}, 
\end{align}
where the following conventions of the one-loop three-point integrals
in $D=4-2\epsilon$ are used
\begin{align}
C_0^{x,y} &= \fr{(\mu_R^2 \pi)^{(4-D)/2}}{i\pi^2}\int d^Dq\fr{1}{D_{x,y}},\\
C_1^{x,y} p_1^\mu+ C_2^{x,y} p_2^\mu&=  \fr{(\mu_R^2 \pi)^{(4-D)/2}}{i\pi^2}\int d^Dq\fr{q^\mu}{D_{x,y}},\\
C_{00}^{x,y} g^{\mu\nu} + C_{11}^{x,y} p_1^\mu p_1^\nu+ C_{12}^{x,y} (p_1^\mu p_2^\nu + p_1^\nu p_2^\mu)+ C_{22}^{x,y} p_2^\mu  p_2^\nu&=  \fr{(\mu_R^2 \pi)^{(4-D)/2}}{i\pi^2}\int d^Dq\fr{q^\mu q^\nu}{D_{x,y}},
\end{align}
where the denominator $D_{x,y}$ is given by 
\be 
D_{x,y}=(q^2- m_x^2)((q-p_1)^2- m_y^2)((q-p_2)^2- m_y^2),
\ee
with $p_1^2= m_l^2, p_2^2=m_l^2$, $(p_1-p_2)^2=0$, $m_l$ being the mass of
the external lepton, $m_{x,y}$ being the masses of the particles $x$
and $y$. 
If $m_l\ll m_x,m_y$ one can use the zero external mass approximation
for these coefficients \cite{Lavoura:2003xp}, 
\begin{align}
C_0(0,0,0,x,y,y)&= \fr{1}{x}\braket{-\fr{1}{t-1} + \fr{\ln(t)}{(t-1)^2}}\,,\\
C_1(0,0,0,x,y,y)&= \fr{1}{x} \braket{ \fr{(t - 3)}{4(t - 1)^2} + \fr{\ln(t)}{2(t - 1)^3} }\,,\\
C_{11}(0,0,0,x,y,y)&=\fr{1}{x} \braket{\fr{(-2 t^2 + 7 t - 11)}{18(t - 1)^3} + 
      \fr{\ln(t)}{3(t - 1)^4} }\,,\\
C_2(0,0,0,x,y,y)&= C_1(0,0,0,x,y,y)\,,\\
C_{22}(0,0,0,x,y,y)&=2 C_{12}(0,0,0,x,y,y)=C_{11}(0,0,0,x,y,y)\,,
\end{align}
where $t= y/x$. If $m_l$ is of the order of the internal masses, one
should use the following expressions 
 \begin{align}
 C_0(m,0,m,x,y,y)&= \int_0^1 \fr{zdz}{-m z^2 + (m+x-y)z - x}\,,\\
 C_1(m,0,m,x,y,y)&= \int_0^1 \fr{-z^2dz}{2(-m z^2 + (m+x-y)z - x)}\,,\\
 C_{11}(m,0,m,x,y,y)&= \int_0^1 \fr{z^3dz}{3(-m z^2 + (m+x-y)z - x)}\,,\\
  C_2(m,0,m,x,y,y)&= C_1(m,0,m,x,y,y)\,,\\
  C_{22}(m,0,m,x,y,y)&=2C_{12}(m,0,m,x,y,y)=C_{11}(m,0,m,x,y,y)\,.
 \end{align}
We have implemented the analytic expressions of these one-loop three-point integral
coefficients 
including the dependence on $m_l^2$ and compared  with the numerical results obtained from the
Package-X \cite{Patel:2015tea}.  The zero external mass approximation  can be
applied for chargino and neutralino one-loop diagrams, and we then  
recover the known formula in the MSSM \cite{Moroi:1995yh}. To the best
of our knowledge, this is the first time that the full one-loop SUSY
corrections to the leptonic AMM and EDM in the complex NMSSM with
inverse seesaw mechanism have been presented. For the  NMSSM without
  inverse seesaw mechanism, the expressions of the full one-loop
  contributions to the muon AMM have been presented in \cite{Domingo:2008bb} and 
  the full one-loop contributions to the electron EDM have been
  discussed in \cite{King:2015oxa}. 
The one-loop chargino and neutralino contributions are always considered to be
dominant in most of the parameter space, so that they are the only ones
taken into account in the analyses of the muon and electron AMM
available in the literature \cite{Cao:2019evo,Cao:2021lmj}. However in case of light sterile
 neutrino masses and/or light singlet-like Higgs bosons, contributions
 from $W$ and/or Higgs diagrams can be significant. Therefore, for the
 investigation of the full parameter space these contributions should
 be taken into account.  
 
%%%%%%%%%%%%%%%%%%%%%%%%%%%%%%%%%%%%%%%%%%%%%%%%%%%%%%%%%%%% 
\subsection{Two-Loop Contributions} 
%%%%%%%%%%%%%%%%%%%%%%%%%%%%%%%%%%%%%%%%%%%%%%%%%%%%%%%%%%%% 
The two-loop SUSY contributions to the muon AMM
in the MSSM have been classified and  evaluated  in \cite{Chen:2001kn,Arhrib:2001xx,Heinemeyer:2003dq,Heinemeyer:2004yq,vonWeitershausen:2010zr,Fargnoli:2013zda,
Fargnoli:2013zia} for the CP-conserving case and in
\cite{Cheung:2009fc} for the CP-violating case. The numerical results of
all two-loop contributions have identified some dominant
contributions. These dominant two-loop SUSY corrections have been
generalized to the CP-conserving NMSSM and implemented in 
\texttt{NMSSMTools} \cite{Domingo:2008bb}. We follow this strategy to
take into account the dominant two-loop contributions. We first
consider the leading-logarithmic two-loop electroweak contribution which
arises from the SUSY one-loop diagrams with an additional photon
loop. This contribution has been evaluated most efficiently by using
the effective Lagrangian approach which can be applied for the 
SM and many new physic models, as perfomed in \cite{Degrassi:1998es}. It is given by 
\be 
a_l^{\text{qed},2l} =-\fr{4\alpha}{\pi} a_l^{\text{new},1l} \log
\fr{M_{\text{SUSY}}}{m_l}, \quad d_l^{\text{qed},2l} =-\fr{4\alpha}{\pi}
d_l^{\text{new},1l} \log \fr{M_{\text{SUSY}}}{m_l},
\ee
where the scale $M_{\text{SUSY}}$ is chosen to be of the order of the
masses of the smuons, in particular $M_{\text{SUSY}}
=\sqrt{m_{\ti \mu_R}m_{\ti \mu_L}} $. 
The negative sign of this term gives a reduction of about ten percent of 
the whole one-loop contribution. 

The Higgs-mediated Barr-Zee-type diagrams \cite{Barr:1990vd} with an internal photon
 can contribute significantly to the leptonic AMM. We consider here the contributions from fermion loops,
 sfermion loops, charged Higgs loops  and chargino loops generating
 the effective $h_i\gamma\gamma $ vertex. These contributions can be
 calculated by evaluating first the effective $h_i\gamma\gamma$ vertex and then
 inserting this effective vertex into the second loop. 
Making use of gauge invariance, the effective  $h_i\gamma\gamma $ vertex 
can be written as\footnote{In the actual calculation, there may appear
  some 
gauge-dependent terms proportional to $q^\mu q^\nu$ and to $g^{\mu\nu}$. They
do not contribute, however, to the EDM and the AMM at two-loop level
as shown in \cite{Abe:2013qla}.} 
\be  
\Gamma_{h_i\gamma\gamma^*}^{\mu\nu} =  ( (k\cdot q)g^{\mu\nu} - q^\mu
k^\nu) \Gamma^A +  \epsilon^{\mu\nu\al\beta}k_\al q_\beta
\Gamma^P, \label{eq:effectiveHyy} 
\ee 
where $k^\mu$, $q^\nu$ are the momenta of the on-shell and off-shell photons, respectively, and $\Gamma^A ,\Gamma^P $ are scalar form factors. 
We evaluate these form factors for  sfermion loops, charged Higgs
loops, chargino loops and fermion loops. They are given by
\begin{align}
\Gamma^A_{\ti f}=& -\fr{N_c^f Q_{\ti f}^2 e^2g_{h_i\ti f^* \ti f}v}{8\pi^2} \int_0^1 dx \fr{x(x-1)}{ q^2x (1-x) - M_{\ti f}^2}  ,\\
\Gamma^P_{\ti f}=&0,\\
\Gamma^A_{H^\pm}=& -\fr{ e^2g_{h_iH^+ H^-}v}{8\pi^2} \int_0^1 dx \fr{x(x-1)}{ q^2x (1-x) - M_{H^\pm}^2} ,\\
\Gamma^P_{H^\pm}=&0,\\
\Gamma^A_{\ti\chi^\pm}=& \fr{ e^3 g^S_{h_i\ti\chi^+_j \ti\chi^-_j} M_{\ti \chi^\pm_j} }{4\pi^2\sqrt{2} s_W} \int_0^1 dx \fr{2x(x-1)+1}{ q^2x (1-x) - M_{\ti \chi^\pm_j}^2},\\
\Gamma^P_{\ti\chi^\pm}=& \fr{ e^3 g^P_{h_i\ti\chi^+_j \ti\chi^-_j} M_{\ti \chi^\pm_j}}{4\pi^2\sqrt{2} s_W} \int_0^1 dx \fr{1}{ q^2x (1-x) - M_{\ti \chi^\pm_j}^2},\\
\Gamma^A_{f}=& \fr{N_c^f Q_{ f}^2 e^2g^S_{h_i \bar f f}m_f^2 }{4\pi^2v} \int_0^1 dx \fr{2x(x-1)+1}{ q^2x (1-x) - m_f^2},\\
\Gamma^P_{f}=& \fr{N_c^f Q_{ f}^2 e^2g^P_{h_i \bar f f}m_f^2 }{4\pi^2v} \int_0^1 dx \fr{1}{ q^2x (1-x) - m_f^2},
\end{align}
where $Q_{f/\ti f}$ is the electric charge of fermion $f$/of sfermion
$\ti f$, $N_c= 3$ for quarks and $N_c= 1$ for leptons. We take into
account only the third generation of quarks and leptons in the loops since they
have significant Yukawa couplings. We used the following convention
for the couplings of the neutral Higgs boson $h_i$ to fermions,
sfermions, charged Higgs bosons and charginos,  
\begin{align}
 &-\fr{i  m_f}{v}\bar f(  g^S_{h_i \bar f f} +i  g^P_{h_i \bar f f}\gamma_5 )fh_i
  - \fr{ig_2}{\sqrt{2}}  \bar \chi^-_j( g^S_{h_i\ti\chi^+_j
   \ti\chi^-_j}  +i g^P_{h_i\ti\chi^+_j \ti\chi^-_j} \gamma_5
   )\chi^-_j h_i \nonumber \\
  &+ iv g_{h_iH^+ H^-}h_iH^+ H^-+ iv g_{h_i\ti f^*_k \ti f_k}h_i\ti f^*_k \ti f_k \,,
\end{align}
with the explicit expressions for $g^{S/P}_{h_i \bar f
  f},\, g^{S/P}_{h_i\ti\chi^+_j \ti\chi^-_j} ,\, g_{h_iH^+ H^-},$ and $g_{h_i\ti
  f^*_k \ti f_k}$  given by 
\begin{align}
g^{S}_{h_i \bar f f}&=\fr{ \mathcal R_{i1}^{H}}{c_\beta} \,, & g^{P}_{h_i \bar f f}&=- \mathcal R_{i4}^{H}t_\beta\,, \quad &f&=b,\tau,e,\mu, \\
g^{S}_{h_i \bar f f}&=\fr{ \mathcal R_{i2}^{H}}{s_\beta} \,, & g^{P}_{h_i \bar f f}&=-\fr{ \mathcal R_{i4}^{H}}
{t_\beta}\,, \quad &f&=t,
\end{align}
%%%%
\vspace*{-0.5cm}
\begin{align}
g^{S}_{h_i\ti\chi^+_j \ti\chi^-_j} &= \Re [U_{j2}^{\chi}V_{j1}^{\chi}\mathcal R_{i1}^{H} + e^{i\varphi_u} U_{j1}^{\chi}V_{j2}^{\chi}
\mathcal R_{i2}^H+ e^{-i\varphi_s}\fr{\lambda^*}{g_2} U_{j2}^{\chi}V_{j2}^{\chi}\mathcal R_{i3}^H] \,\crn
& - \Im[s_\beta U_{j2}^{\chi}V_{j1}^{\chi} + e^{i\varphi_u} c_\beta U_{j1}^{\chi}V_{j2}^{\chi} ]\mathcal R_{i4}^H 
+  \Im[\fr{\lambda^* e^{-i\varphi_s}}{g_2} U_{j2}V_{j2}  ]\mathcal R_{i5}^H  \,,\\
g^{P}_{h_i\ti\chi^+_j \ti\chi^-_j} &= \Im [U_{j2}^{\chi}V_{j1}^{\chi}\mathcal R_{i1}^H + e^{i\varphi_u} U_{j1}^{\chi}V_{j2}^{\chi}
\mathcal R_{i2}^H+ e^{-i\varphi_s}\fr{\lambda^*}{g_2} U_{j2}^{\chi}V_{j2}^{\chi}\mathcal R_{i3}^H] \,\crn
& + \Re[s_\beta U_{j2}^{\chi}V_{j1}^{\chi} + e^{i\varphi_u} c_\beta U_{j1}^{\chi}V_{j2}^{\chi} ]\mathcal R_{i4}^H 
-  \Re[\fr{\lambda^* e^{-i\varphi_s}}{g_2} U_{j2}^{\chi}V_{j2}^{\chi}  ]\mathcal R_{i5}^H  \,,\\
g_{h_iH^+ H^-}&= \left(\fr{g_1^2 c_\beta c_{2\beta}}{4} -\fr{g_2^2 c_\beta (1+2s_\beta^2)}{4}
+c_\beta s_\beta^2 |\lambda|^2 \right)\mathcal R_{i1}^H\,\crn
&+\left(-\fr{g_1^2 s_\beta c_{2\beta}}{4} -\fr{g_2^2 s_\beta (1+2c_\beta^2)}{4}
+c_\beta^2 s_\beta |\lambda|^2 \right)\mathcal R_{i2}^H\,\crn
&+\bigg(-\fr{ s_{2\beta}\Re[A_\lambda \lambda e^{i(\varphi_u+\varphi_s)}] }{\sqrt{2} v} 
-\fr{ s_{2\beta}v_s\Re[\kappa^*\lambda e^{i(\varphi_u-\varphi_s)}] }{ v} -\fr{|\lambda|^2v_s}{v}\bigg)
\mathcal R_{i3}^H\,\crn
& +\bigg(\fr{ s_{2\beta}\Im[A_\lambda \lambda e^{i(\varphi_u+\varphi_s)}] }{\sqrt{2} v} 
-\fr{ s_{2\beta}v_s\Im[\kappa^*\lambda e^{i(\varphi_u-\varphi_s)}] }{ v} \bigg)
\mathcal R_{i5}^H\,,
\end{align}
%%%%
\begin{align}
 g_{h_i\ti t^*_k \ti t_k}&= \bigg(  \fr{g_1^2}{12}c_\beta (|U^{\ti t}_{k1}|^2-
4 |U^{\ti t}_{k2}|^2
 )  -\fr{g_2^2}{4}c_\beta|U^{\ti t}_{k1}|^2+ \fr{y_tv_s\Re[\lambda e^{I\varphi_s} 
 U^{\ti t*}_{k1}U^{\ti t}_{k2}] }{v} \bigg)\mathcal R_{i1}^H\,\crn
 &+ \bigg(  \fr{g_1^2}{12}s_\beta (-|U^{\ti t}_{k1}|^2+
4 |U^{\ti t}_{k2}|^2
 )  +\fr{g_2^2}{4}s_\beta|U^{\ti t}_{k1}|^2+ y_t^2s_\beta 
 (|U^{\ti t}_{k1}|^2 +|U^{\ti t}_{k2}|^2)  \crn
 %%%
 &-\fr{\sqrt 2 y_t \Re[A_t e^{i\varphi_u} U^{\ti t*}_{k1}U^{\ti t}_{k2}] }{v} \bigg)\mathcal R_{i2}^H
 +\Re[ \lambda e^{i\varphi_s}  U^{\ti t}_{k1}U^{\ti t*}_{k2}]y_tc_\beta  \mathcal R_{i3}^H\,
 \crn
 &+\left(- \fr{\Im[\lambda e^{i\varphi_s} U^{\ti t}_{k1}U^{\ti t*}_{k2} ]y_tv_s s_\beta }{v} + \fr{\sqrt{2}c_\beta y_t \Im[A_f e^{i\varphi_u} 
 U^{\ti t*}_{k1}U^{\ti t}_{k2}] }{v}\right)\mathcal R_{i4}^H\,\crn
 &- c_\beta  y_t \Im[\lambda e^{i\varphi_s } U^{\ti t}_{k1}U^{\ti t*}_{k2} ] \mathcal R_{i5}^H\,,
\end{align}
%%%%%%
\begin{align}
g_{h_i\ti b^*_k \ti b_k}&= \bigg(  \fr{g_1^2}{12}c_\beta (|U^{\ti b}_{k1}|^2+
4 |U^{\ti b}_{k2}|^2
 )  +\fr{g_2^2}{4}c_\beta|U^{\ti b}_{k1}|^2- y_b^2c_\beta ( U^{\ti b*}_{k1}U^{\ti b}_{k1}
 +U^{\ti b*}_{k2}U^{\ti b}_{k2}) \,\crn
 &
  - \fr{\sqrt2 y_b \Re[A_bU^{\ti b*}_{k1}U^{\ti b}_{k2} ]}{v} \bigg)\mathcal R_{i1}^H+ \bigg(  \fr{g_1^2}{12}s_\beta (-|U^{\ti b}_{k1}|^2-2
 |U^{\ti b}_{k2}|^2
 ) \,\crn
 & -\fr{g_2^2}{4}s_\beta|U^{\ti t}_{k1}|^2+
 \fr{ y_b v_s\Re[ \lambda e^{i(\varphi_u+\varphi_s)} U^{\ti b}_{k1}U^{\ti b*}_{k2}] }{v} \bigg)\mathcal R_{i2}^H+\Re[ \lambda e^{i(\varphi_s+\varphi_u)}  U^{\ti b}_{k1}U^{\ti b*}_{k2}]y_bs_\beta  \mathcal R_{i3}^H
 \,\crn
 &
 +\left(- \fr{\Im[\lambda e^{i(\varphi_s+\varphi_u)} U^{\ti b}_{k1}U^{\ti b*}_{k2} ]y_tv_s c_\beta }{v} + \fr{\sqrt{2}s_\beta y_b \Im[A_b  
 U^{\ti b*}_{k1}U^{\ti b}_{k2}] }{v}\right)\mathcal R_{i4}^H\,\crn
 &- s_\beta  y_b \Im[\lambda e^{i(\varphi_s+\varphi_u) } U^{\ti b}_{k1}U^{\ti b*}_{k2} ] \mathcal R_{i5}^H\,,
\end{align}
%%%
%%%%%%
\begin{align}
g_{h_i\ti \tau^*_k \ti \tau_k}&= \bigg(  \fr{g_1^2}{4}c_\beta (|U^{\ti \tau}_{k1}|^2-
2 |U^{\ti \tau}_{k2}|^2
 )  +\fr{g_2^2}{4}c_\beta|U^{\ti \tau}_{k1}|^2- y_\tau^2c_\beta ( U^{\ti \tau*}_{k1}U^{\ti \tau}_{k1}
 +U^{\ti \tau*}_{k2}U^{\ti \tau}_{k2})  \,\crn
 &
 - \fr{\sqrt2 y_\tau \Re[A_\tau U^{\ti \tau*}_{k1}U^{\ti \tau}_{k2} ]}{v} \bigg)\mathcal R_{i1}^H+ \bigg(  \fr{g_1^2}{4}s_\beta (|U^{\ti \tau}_{k1}|^2-2
 |U^{\ti \tau}_{k2}|^2
 ) \,\crn
 & -\fr{g_2^2}{4}s_\beta|U^{\ti \tau}_{k1}|^2+
 \fr{ y_\tau v_s\Re[ \lambda e^{i(\varphi_u+\varphi_s)} U^{\ti \tau}_{k1}U^{\ti \tau*}_{k2}] }{v} \bigg)\mathcal R_{i2}^H
 \,\crn
 &
 +\left(- \fr{\Im[\lambda e^{i(\varphi_s+\varphi_u)} U^{\ti \tau}_{k1}U^{\ti \tau*}_{k2} ]y_\tau v_s c_\beta }{v} + \fr{\sqrt{2}s_\beta y_\tau \Im[A_\tau  
 U^{\ti \tau*}_{k1}U^{\ti \tau}_{k2}] }{v}\right)\mathcal R_{i4}^H\,\crn
 &+\Re[ \lambda e^{i(\varphi_s+\varphi_u)}  U^{\ti \tau}_{k1}U^{\ti \tau*}_{k2}]y_\tau s_\beta  \mathcal R_{i3}^H\, - s_\beta  y_\tau \Im[\lambda e^{i(\varphi_s+\varphi_u) } U^{\ti \tau}_{k1}U^{\ti \tau*}_{k2} ] \mathcal R_{i5}^H.
\end{align}
%%%

 The gauge invariant form 
of the effective $h_i\gamma\gamma$ coupling in \eqref{eq:effectiveHyy}
will be inserted into the second loop to get the AMM and EDM. The
lepton mass in the numerator of the second loop is neglected, since
this leads to contributions suppressed by the factor $m_l^2/M_S^2$
where $M_S$ is the mass of the heavy particles.  
We present here the analytic expressions for the AMM from sfermion
loops, charged Higgs loops, chargino loops and fermion loops, 
\begin{align}
a_l^{ \ti f,2l} =& \sum_{i=1}^5 \fr{N_c^f Q_{\ti f}^2 \alpha m^2_l}{16\pi^3m_{h_i}^2} (g^S_{h_i \bar ll} g_{h_i\ti f^* \ti f}) {\cal F}^{(2)}\left( \fr{M_{\ti f}^2}
{m_{h_i}^2}\right),\\
a_l^{H^\pm,2l} =& \sum_{i=1}^5 \fr{ \alpha m^2_l}{16\pi^3m_{h_i}^2} (g^S_{h_i \bar ll} g_{h_iH^+ H^-}) {\cal F}^{(2)}\left( \fr{M_{H^\pm}^2}
{m_{h_i}^2}\right),\\
a_l^{\ti\chi^\pm,2l } =& \sum_{i=1}^5 \sum_{j=1}^2\fr{ \alpha^2 m^2_l}{2 \sqrt{2} \pi^2 M_W s_W^2} \fr{1}{M_{\ti \chi^\pm_j}}\left(g^S_{h_i \bar ll} g^S_{h_i\ti\chi^+_j \ti\chi^-_j}{\cal F}^{(1)}\left( \fr{M_{\ti \chi^\pm_j}^2}
{m_{h_i}^2} \right)-g^P_{h_i \bar ll} g^P_{h_i\ti\chi^+_j \ti\chi^-_j} {\cal G}\left( \fr{M_{\ti \chi^\pm_j}^2}
{m_{h_i}^2} \right) \right) ,\\
a_l^{ f,2l} =& \sum_{i=1}^5 \fr{N_c^f Q_{\ti f}^2 \alpha^2 m^2_l}{4 \pi^2 M_W^2 s_W^2} \left(g^S_{h_i \bar ll} g^S_{h_i \bar f f}{\cal F}^{(1)}\left( \fr{m_{f}^2}
{m_{h_i}^2} \right)- g^P_{h_i \bar ll} g^P_{h_i\bar f f} {\cal G}\left( \fr{m_{f}^2}
{m_{h_i}^2} \right) \right) ,
\end{align} 
where $\alpha$ is the fine structure constant and the two-loop functions are given by
\begin{align}
{\cal F}^{(2)}(z)&=  \int_0^1 dx \fr{x(1-x)}{z+ x(x-1)} \ln
                   \fr{x(1-x)}{z} \,,\\
{\cal F}^{(1)}(z)&= \fr{z}{2} \int_0^1 dx \fr{1-2x(1-x)}{x(1-x)-z} \ln
                   \fr{x(1-x)}{z} \, ,\\
{\cal G}(z)&= \fr{z}{2} \int_0^1 dx \fr{1}{x(1-x)-z} \ln
             \fr{x(1-x)}{z} \, .
\end{align}
Note that these expressions  are in agreement  with Eq.~(3.8) of
Ref.~\cite{Cheung:2009fc} for the complex MSSM, we used, however, a
different sign convention compared to their notation. These two-loop contributions 
are then subtracted from the corresponding SM contributions arising
from top, bottom quark and tau lepton loops. 
The leptonic EDM can be obtained from the above formulae with the replacement
\be 
d_l^{ x,2l} =  \fr{e}{2m_l}a_l^{x,2l}( g^S_{h_i \bar ll} \to g^P_{h_i
  \bar ll}, g^P_{h_i \bar ll}\to -g^S_{h_i \bar ll}), \quad x= \ti f,
H^\pm, \ti\chi^\pm,  f.  
\ee
The two-loop Barr-Zee-type contributions to the electron EDM have been
implemented in {\tt NMSSMCALC} as
described in Ref.~\cite{King:2015oxa}. It 
does not only contain contributions coming from the effective
$h_i\gamma\gamma $ vertex but also other contributions arising from
the effective $h_i\gamma Z $, $H^\pm\gamma W^\mp $, $H^\pm\gamma
W^\mp$ vertices. Since there is no difference between the two models 
in these contributions we keep them unchanged  in {\tt NMSSMCAL-nuSS}. 

In summary, the SUSY contributions to the leptonic AMM and EDM
considered in this study are the sum of the full one-loop and partial
two-loop contributions, 
\bea
a_l &= a_l^{1l}+ a_l^{\text{qed},2l}+a_l^{ \ti f,2l}+a_l^{H^\pm,2l}+a_l^{\ti\chi^\pm,2l }+(a_l^{ f,2l}-a_l^{\text{SM},f,2l})\,\\
d_l &= d_l^{1l}+ d_l^{\text{qed},2l}+d_l^{ \ti f,2l}+d_l^{H^\pm,2l}+d_l^{\ti\chi^\pm,2l }+(a_l^{ f,2l}-a_l^{\text{SM},f,2l})\,\crn
&  + d_l^{2l}(h_i\gamma Z)+d_l^{2l}(H^\pm\gamma W^\mp)+d_l^{2l}(H^\pm\gamma
W^\mp)
\eea 
where $d_l^{2l}(h_i\gamma Z)+d_l^{2l}(H^\pm\gamma W^\mp)+d_l^{2l}(H^\pm\gamma
W^\mp)$ are the two-loop Barr-Zee-type contributions arising from
the effective $h_i\gamma Z $, $H^\pm\gamma W^\mp $, $H^\pm\gamma
W^\mp$ vertices.

%%%%%%%%%%%%%%%%%%%%%%%%%%%%%%%%%%%%%%%%%%%%%%%%%%%%%%%%%%%%%%
%%%%%%%%%%%%%%%%%%%%%%%%%%%%%%%%%%%%%%%%%%%%%%%%%%%%%%%%%%%%%%
\section{Numerical Analysis \label{sec:results}}
%%%%%%%%%%%%%%%%%%%%%%%%%%%%%%%%%%%%%%%%%%%%%%%%%%%%%%%%%%%%%%
In this section we investigate the numerical impact of the
neutrino/sneutrino sector  and  
various CP-violating phases on the muon AMM 
 and on the electron EDM. 
It has been shown in our study in \cite{Dao:2021vqp}, that the extended neutrino 
and sneutrino sectors can have a significant impact on the Higgs
sector, the charged lepton flavor-violating decays, $l_i\to l_j+\gamma$, and  
the new physics constraints from the oblique parameters $S, T,
U$. We therefore will investigate also
  what is the correlation between these impacts.
In order to find viable parameter points we performed a 
scan in the NMSSM parameter space. 
We have used {\tt NMSSMCALC-nuSS} to calculate 
the Higgs boson masses including the available two-loop
corrections at ${\cal O}(\al_s\al_t +\al_t^2)$,\footnote{Note that we
  have taken into account the complete one-loop corrections computed in the
  NMSSM with inverse seesaw mechanism \cite{Dao:2021vqp}, but took
  over the two-loop corrections from the pure NMSSM.} the Higgs decay widths and branching
ratios including the state-of-the-art higher-order QCD corrections
as well as the Higgs effective couplings. We then use {\tt
  HiggsBounds} \cite{Bechtle:2020pkv} to check if the parameter points
pass all the exclusion limits from the searches at
LEP, Tevatron and the LHC, and {\tt HiggsSignals-2.6.1}
\cite{Bechtle:2020uwn} to check if the points are consistent
with the LHC data for a 125 GeV Higgs boson. A parameter point is chosen if it is 
consistent with the Higgs data within 2$\sigma$. With our {\tt
  NMSSMCALC-nuSS} code we can also 
check if the parameter point is in accordance with the active light neutrino data,
the constraints from the charged lepton flavor-violating decays and
the electroweak observables, see \cite{Dao:2021vqp} for more information. 
 
In order to show the impact of the neutrino Yukawa couplings on the AMM 
we choose a sample parameter point from our generated scan sample
satisfying all the mentioned constraints, called {\tt P1} in the
following. The SM input parameters are taken from the Particle Data
Group \cite{Zyla:2020zbs} and are given by
\begin{equation}
\begin{tabular}{lcllcl}
\quad $\alpha(M_Z)$ &=& 1/127.955, &\quad $\alpha^{\overline{\mbox{MS}}}_s(M_Z)$ &=&
0.1181\,, \\
\quad $M_Z$ &=& 91.1876~GeV\,, &\quad $M_W$ &=& 80.379~GeV \,, \\
\quad $m_t$ &=& 172.74~GeV\,, &\quad $m^{\overline{\mbox{MS}}}_b(m_b^{\overline{\mbox{MS}}})$ &=& 4.18~GeV\,, \\
\quad $m_c$ &=& 1.274~GeV\,, &\quad $m_s$ &=& 95.0~MeV\,,\\
\quad $m_u$ &=& 2.2~MeV\,, &\quad $m_d$ &=& 4.7~MeV\,, \\
\quad $m_\tau$ &=& 1.77682~GeV\,, &\quad $m_\mu$ &=& 105.6584~MeV\,,  \\
\quad $m_e$ &=& 510.9989~keV\,, &\quad $G_F$ &=& $1.16637 \cdot 10^{-5}$~GeV$^{-2}$\,.
%\label{eq:param1} 
\end{tabular}
\end{equation} The light neutrino input parameters 
are set equal to their best-fit
values \cite{Zyla:2020zbs} together with a fixed value for the lightest neutrino mass, in particular,
\begin{table}[h]
\centering
\begin{minipage}[t]{0.5\textwidth}
\centering
    \begin{tabular}[t]{ll}
 $m_{\nu_1}$ &= $10^{-11} \,\gev\,,$ \\
 $m_{\nu_2} $&=$  \sqrt{ m_{\nu_1}^2 + 7.37\times 10^{-23}}\, \gev\,,$\\
$ m_{\nu_3}$ &=$ \sqrt{ m_{\nu_1}^2 + 2.525\times 10^{-21}}\, \gev\,,$\\
\end{tabular}
\end{minipage}\hfill
\begin{minipage}[t]{0.5\textwidth}
\centering
    \begin{tabular}[t]{ll}
$\theta_{12} $&=$  \arcsin(\sqrt{0.297})\,,$\\
$\theta_{23} $&=$  \arcsin(\sqrt{0.425})\,,$\\
$\theta_{13} $&=$  \arcsin(\sqrt{0.0215})\,,$\\
$\delta_{CP} $&=$  248.4^\circ\,.$\\
\end{tabular}
\end{minipage}
\label{bestfitp}
\end{table}

All other complex phases are set to zero and
the remaining input parameters are given by 
\begin{align}
M_{H^\pm}            &= 1000   \, \gev \,,   &  m_{\tilde{\mu}_L}&=400\,\gev\,,          \crn
M_1             & =400  \, \gev  \,,      &  m_{\tilde{\mu}_R}&=m_{\tilde{q}_L}=m_{\tilde{q}_R}=2000\,\gev\,,       \crn    
M_2              & =400  \, \gev   \,,   &  m_{\ti X}            & =0\, \gev \,,     \crn             
\mueff              & =400 \, \gev    \,, &  m_{\ti N}           & =0 \, \gev  \,,      \crn      
m_{\tilde{Q}_3} & = 1000 \, \gev \,, &  A^{\nu}_{11}=A^{\nu}_{33}              & =1000 \, \gev \,,       \\
 m_{\tilde{t}_R}     & =1500 \, \gev \,,  &   A^{\nu}_{22}              & =-1000 \, \gev \,,   \crn   
 m_{\tilde{e}_L}= m_{\tilde{\tau}_L}     & =2000 \, \gev \,,  & A_{X}              & =1000 \, \gev  \,,     \crn             
m_{\tilde{e}_R}=m_{\tilde{\tau}_R}      &=2000 \, \gev \,, &  \mu_X              & =600 \, \gev \,,       \crn
A_t                  & =2000  \, \gev \,,  &          B_{\mu_X}              & =10 \, \gev  \,,                   \crn        
\mbox{Re} A_k             & =-100 \, \gev   \,,   &  y^\nu_{11}&=y^\nu_{33} =0.5\,,\quad i=1,2,3\,,               \crn        
\tan\beta           & =12  \,,      &  y^\nu_{22}&=0.9  \,,       
                                                                      \crn     
\lambda             &= 0.252   \,,      &   \kappa    &=0.297  \,.          \nonumber
\end{align}
Note that we have used the $\mu_X$-parameterization  where the
neutrino Yukawa couplings $y^\nu$ are given as inputs. For the parameter point {\tt P1}, we have chosen  $y^\nu$ to be a diagonal matrix.
With this choice, we do not need to worry about the violation of the charged lepton flavor-violating decays, $l_i\to l_j+\gamma$. 
In \tab{tab:massP1OS}, we present the Higgs mass spectrum  with and
without inverse seesaw mechanism at two-loop $\orderO{\alpha_t \alpha_s +
  \alpha_t^2}$ using the OS renormalization for the top/stop
sector. For the parameter point {\tt P1}, the stop masses 
  in the OS scheme are given by
  \be m_{\ti t_1}= 1001.62\, \gev\,,\quad m_{\ti t_2}= 1524.9\, \gev\,.\ee  The  main components of the Higgs mass eigenstates are
also shown in the last row.
\begin{table}[h]
\begin{center}
\begin{tabular}{|l|l|c|c|c|c|c|}
\hline
                              \multicolumn{2}{|l|}{}         & ${h_1}$   & ${h_2}$ & ${h_3}$ & ${h_4}$ & ${h_5}$ \\ \hline \hline
\multirow{2}{*}{${\cal O}(\alpha_t \alpha_s+ \alpha_t^2)$} & without ISS &124.2 & 369.59 & 912.37& 998.91   & 999.94  \\
             & with ISS                      &125.46 & 369.65 & 912.40& 998.85 & 1000.0     \\ \hline
\multicolumn{2}{|l|}{main component}                                    & $h_u$     & $a_s$   & $h_d$   & $a$   & $h_s$   \\ \hline
\end{tabular}
\caption{Parameter point {\tt P1}:  Mass values in GeV and main
  components of the neutral Higgs bosons at 
  two-loop $\orderO{\alpha_t \alpha_s +  \alpha_t^2}$ obtained for the 
  NMSSM without and with the inverse seesaw mechanism using $\OS$
  renormalization in the top/stop sector. }
\label{tab:massP1OS}
\end{center}
\end{table}
As can be inferred from \tab{tab:massP1OS}, the neutrino/sneutrino
sector increases the loop correction to the SM-like Higgs boson given
by the $h_u$-like state.
The spectrum of the electroweakinos  and smuons is the same in both
models and is given in \tab{tab:weakinoP1OS}.  
\begin{table}[h]
\begin{center}
\begin{tabular}{|c|c|c|c|c|c|c|c|c|}
\hline
                           ${\ti \chi_1^0}$   
                              & ${\ti \chi_2^0}$ & ${\ti \chi_3^0}$ & ${\ti \chi_4^0}$ 
                              & ${\ti \chi_5^0}$
                              & ${\ti \chi_1^+}$& ${\ti \chi_2^+}$ & $\ti \mu_1$ & $\ti \mu_2$\\ \hline \hline
331.93 & 400.00 & 405.14 & 470.96   & 945.1& 341.19 & 461.72 &402.83 & 2000   
              \\ \hline
%  $\ti W^0$     & $\ti B$   & $\ti h_d$   & $\ti h_u$   & $\ti h_s$&  $\ti W^+$ & $\ti H^+_u$
%  & $\ti \mu_L$& $\ti \mu_R$\\ \hline
\end{tabular}
\caption{Electroweakino and smuon masses in GeV for the parameter
  point {\tt P1}. }
\label{tab:weakinoP1OS}
\end{center}
\end{table} 
In the sneutrino sector, the left-handed muon dominated sneutrino mass
is about $394.84\,\gev$ in the NMSSM without inverse seesaw
mechanism. In the NMSSM-nuSS, the muon-like sneutrino is the
  lightest superparticle (LSP) and has a mass of $261.6\,\gev$. In the
  NMSSM without inverse seesaw mechanism the LSP is given by the wino-like
  neutralino.
%Note that
%we have taken into account the rough exclusion limit on the search for charged slepton and 
%chargino in which $m_{\ti l} <159\,\gev$ and $m_{\chi_1^\pm}<135\,\gev$ are excluded at 95\%\,CL 
%for the case mass splitting between the chargino and the LSP up to 100 GeV 
%\cite{ATLAS:CONF:2022:006}.

\paragraph{Impact on the muon AMM:}
In \tab{tab:ammNMSSM} we present for the NMSSM  without inverse
  seesaw mechanism the individual contributions to the muon AMM as
  well as its total value. If not stated otherwise 
the results of the AMM of the muon are normalized to $10^{-10}$. 
\begin{table}[h]
 \renewcommand{\arraystretch}{1.3}
\begin{center}
\setlength\tabcolsep{0.05cm}
\begin{tabular}{|c|c|c|c|c|c|c|c|c|c|}
\hline
                           $a_\mu^{\ti \chi^0,1l}$   
                              & $a_\mu^{\ti \chi^\pm,1l}$& $a_\mu^{H,1l}$ & $a_\mu^{H^\pm,1l}$ &$a_\mu^{\text{qed},2l}$
                              & $a_\mu^{\ti f,2l}$
                              & $a_\mu^{f,2l}$ & $a_\mu^{ H^\pm,2l}$& $a_\mu^{ \ti \chi^\pm,2l}$& $a_\mu$ \\ \hline \hline
$-1.30 $& $11.55 $& $2\times 10^{-5}$ & $-6\times 10^{-6}$ & $-2.90 $ & $3\times 10^{-3}$  & $-3
\times 10^{-2}$ &$ 3\times 10^{-4}$ &$-6\times 10^{-2}$ & $7.26$   
              \\ \hline
\end{tabular}
\caption{The individual contributions to the muon AMM  in the
  NMSSM without inverse seesaw mechanism. The
  total sum is given in the last column. All values are
    normalized to $10^{-10}$.}
\label{tab:ammNMSSM}
\end{center}
\end{table} 
The dominant contribution comes from the chargino one-loop
diagram. The contributions from the neutral Higgs and charged Higgs
one-loop diagrams  are very small since they are both proportional to
$m_\mu^4$. The second and third important contributions are  the
two-loop SUSY QED and  the neutralino one-loop ones. They are both
negative. The other two-loop contributions are small and negligible
for the parameter point {\tt P1} where the masses of the
non-SM-like Higgs bosons and the SUSY paticles are rather heavy.  

In the NMSSM-nuSS, the neutrino/sneutrino sector significantly changes
the  one-loop and the two-loop QED contributions while the two-loop
contributions including the $h_i\gamma\gamma$ effective couplings
remain unchanged w.r.t.~the pure NMSSM. We present in
\tab{tab:ammNMSSM-nuSS} the individual contributions from the one-loop
diagrams as well as the two-loop QED contribution to the AMM of the muon
  and its total sum. 
\begin{table}[h]
\begin{center}
\begin{tabular}{|c|c|c|c|c|c|c|c|c|c|}
\hline
                          $a_\mu^{\ti \chi^0,1l}$   
                              & $a_\mu^{\ti \chi^\pm,1l}$& $a_\mu^{H,1l}$ & $a_\mu^{H^\pm,1l}$   & $a_\mu^{W,1l}$  &$a_\mu^{\text{qed},2l}$
                            & $a_\mu$ \\ \hline \hline
$-1.3 $& $26.99 $& $2\times 10^{-5}$&  $-0.286$ & $-1.3$  & $-6.82 $& $17.21$   
              \\ \hline
\end{tabular}
\caption{The individual one-loop and two-loop QED contributions to the
  AMM of the muon in the NMSSM-nuSS. The sum of all contributions is
  presented in the last column. All values are
    normalized to $10^{-10}$.}
\label{tab:ammNMSSM-nuSS}
\end{center}
\end{table} 
With the light sneutrino masses and large muon-neutrino Yukawa
coupling $y_{22}^{\nu}$, the chargino one-loop contribution has
increased by a factor of about 2.3 compared to that of the NMMSM without ISS. The same behavior has been observed in Ref.~\cite{Cao:2019evo}. This can be 
seen explicitly from the coupling $g^R_{l\ti\chi_j^- \ti\nu_i} $ of
the chargino with the muon and the sneutrino 
presented in \eqref{eq:charginoCL} where the second term is
propotional to the neutrino 
Yukawa coupling $y_{ll}^\nu$. Depending on the relative sign between
the first and the second term in $g^R_{l^+
\ti\chi_j^- \ti\nu_i} $, as well as on the sneutrino spectrum, the
sneutrino contribution can increase or 
decrease the one-loop chargino contribution. A surprisingly large
  change is also observed in the $W$-boson and charged Higgs one-loop
contributions. Note that we subtract the $W$-boson SM contribution
  from the $W$-boson contribution in the NMSSM-nuSS as mentioned in 
\ssect{sec:one-loopAMM}. In the NMSSM without ISS, the $W$-boson
contribution is exactly equal to the SM one, that is why 
it does not appear in \tab{tab:ammNMSSM}. To understand better the $W$-boson
contribution in the NMSSM-nuSS, we look at the neutrino spectrum. For
this particular point $\mu_X$ has been set to $600\,\gev$, so that
there are four sterile neutrinos, two with a mass of about $
600\,\gev$ and two with mass around $619\,\gev$. We have
tried to reduce $\mu_X$  to decrease the sterile  
neutrino masses so that the magnitude of the $W$-boson contribution
increases. But this also leads to the violation of the unitarity
constraint, see \cite{Dao:2021vqp} for the definition of this
constraint. The magnitude of the charged Higgs contribution has  
 increased by a factor of about $10^5$ compared to the NMSSM without
 ISS. This is because in the NMSSM without ISS, the charged
   Higgs contribution is suppressed by the factor $m_l^4/v^2$ while in
   the model with ISS there appears a new contribution being
   proportional to $m_l^2 y_{ll}^\nu$, see \eqref{eq:chargedHiggsCT}. This
 contribution can be ${\cal O}(10^{-10})$ if the
  charged Higgs mass is light enough. 
    For our parameter point, the charged Higgs mass is 1~TeV, so that
    its contribution is of ${\cal O}(10^{-11})$ 
which does not play an important role in the sum of all contributions.
 
\begin{figure}[h]
    \centering
        \includegraphics[width=0.5\textwidth]{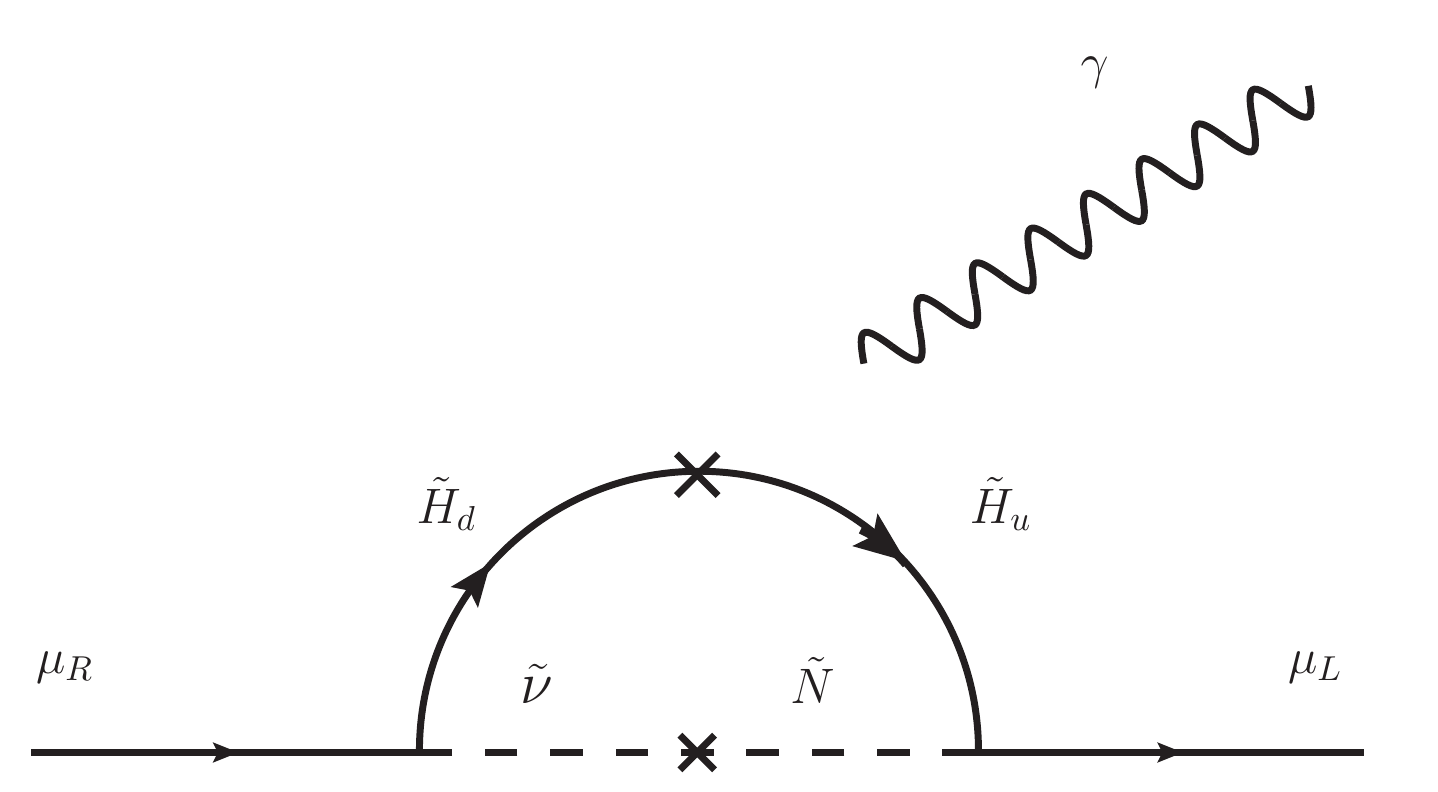}
    \caption{The new enhancement mechanism contributing to the
      anomalous magnetic moment of the muon in the NMSSM-nuSS. 
  } 
    \label{fig:OL-AMM-EN}
\end{figure}

\paragraph{Comparison with the impact on the SM-like Higgs mass:}
 We now investigate the impact of the neutrino and sneutrino parameters on
 the muon AMM in the NMSSM-nuSS in comparison to their impact
 on the SM-like Higgs boson mass. Starting from the parameter 
 point {\tt P1}, we have varied several parameters to see the change
 of  the sum of all contributions $a_\mu$ to the AMM. We can divide
 them into two sets. The first set contains parameters that change the
 muon-neutrino Yukawa coupling $y_{22}^{\nu}$. It enters directly 
   the couplings of the $W$ boson, the charged Higgs and the chargino
   with neutrinos.  
In the $\mu_X$ parameterization, it is $y_{22}^{\nu}$ that is
  changed, while in the Casas-Ibarra parameterization it is  
  $\mu_X$ and $\lambda_X$ that are changed. The second set includes
parameters that result in a significant change of the spectrum of the
sneutrino masses. The sneutrino trilinear coupling $A^\nu_{22},
A^X_{22}$,  and the soft SUSY-breaking masses $\ti m^{X}_{22},\ti
m^{N}_{22}, B^{\mu_X}_{22}$ belong to the second set.  
%
%%%%%%%%%%
\begin{figure}[h]
    \centering
    \subfloat[]{
    \begin{tabular}[b]{r}
        \includegraphics[width=0.495\textwidth]{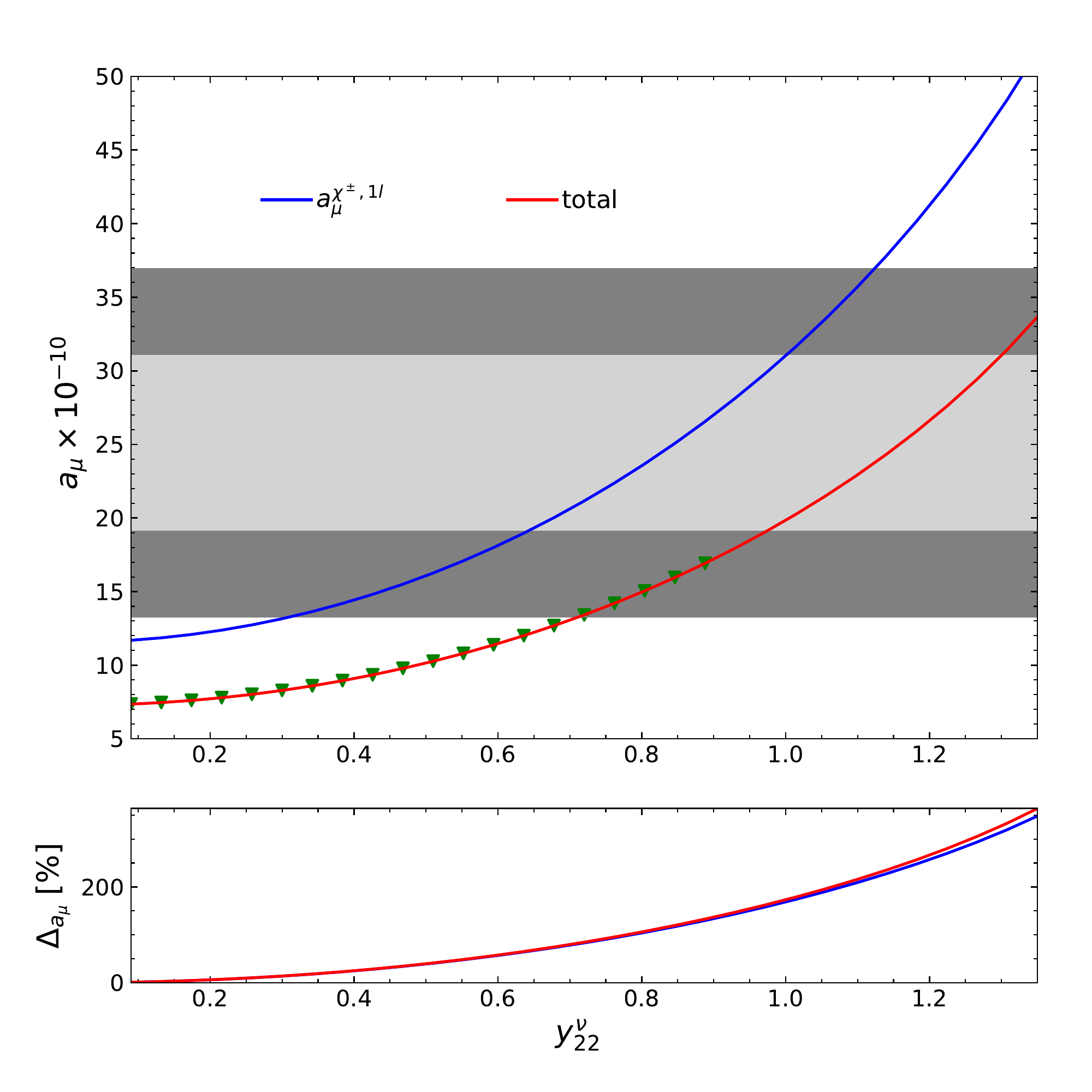}
    \end{tabular}
    }
    \subfloat[]{
    \begin{tabular}[b]{r}
        \includegraphics[width=0.495\textwidth]{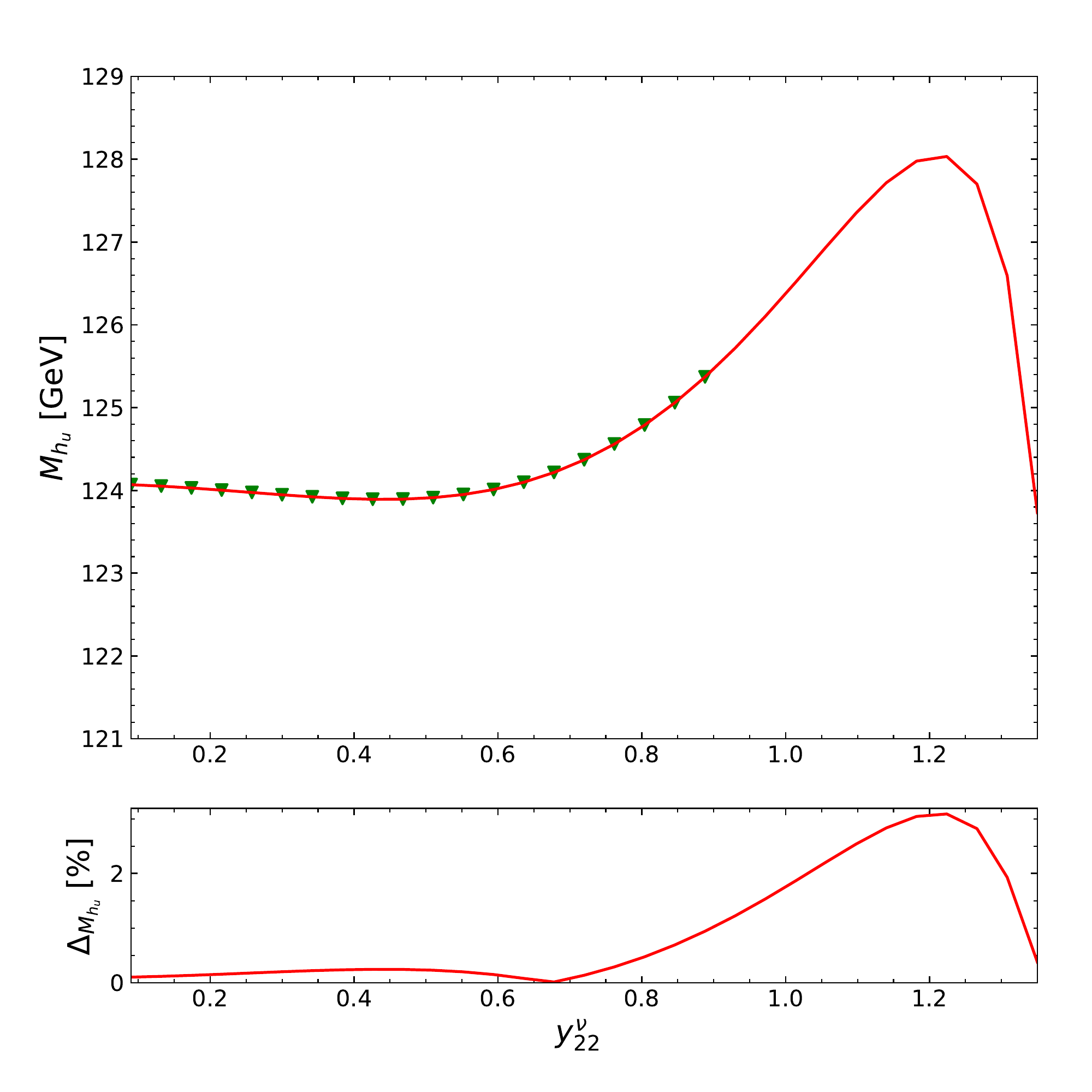}
    \end{tabular}
    }
    \caption{Upper left: The one-loop chargino contribution (blue)  and the
      sum of all contributions (red)  to the AMM of the muon in the
      NMSSM-nuSS. Lower left: The corresponding relative
      difference defined as $\Delta_{a_\mu} =
      \big|(a^{x}_\mu(\text{NMSSM-nuSS})- a^{x}_\mu(\text{NMSSM})
      )/a^{x}_\mu(\text{NMSSM})\big|$. Both as a function of  $y_{22}^{\nu}$ in the $M_X$
      parameterization. Upper right: 
The loop-corrected mass of the $h_u$-like Higgs boson in GeV at order $\orderO{\alpha_t
  \alpha_s +  \alpha_t^2}$. Lower right: The corresponding relative
difference $\Delta_{M_{h_u}}= \left|(M_{h_u}^{\text{NMSSM-nuSS}}
  -M_{h_u}^{\text{NMSSM}} )/M_{h_u}^{\text{\text{NMSSM}}}
\right|$. Both as a function of $y_{22}^{\nu}$ in the $\mu_X$
      parameterization.}  
    \label{fig:ynu}
\end{figure}
%%%%%%%%%%  
In  \figref{fig:ynu} we vary $y_{22}^{\nu}$ in the range $[0.9,1.35]$, keeping the other parameters fixed as 
in  the parameter point {\tt P1} using the $\mu_X$ parameterization. If
$y_{22}^{\nu}>1.45$, one enters the region where the sneutrino mass
squared becomes negative. We remind the reader that $y_{22}^{\nu}$
enters the couplings between the muon, the charginos and the sneutrinos and also
enters the mass matrix of the sneutrinos. 
 Increasing  $y_{22}^{\nu}$ leads to an increase of the mixing between
 the left-handed muon sneutrinos $\ti \nu_2$,  $\tN_2$, $\tX_2$,
 so that the mass of the $\ti \nu_2$-like sneutrino becomes smaller
 while the mass of the $\tN_2$-like sneutrino increases. In the upper
 left plot, we  show the dependence of the  one-loop chargino
 contribution (blue) and the sum of all contributions (red)  to the  
AMM of the muon in the NMSSM-nuSS as a function of $y_{22}^{\nu}$. We see
a strong dependence on $y_{22}^{\nu}$ which can be understood by using
an approximate expression for the new contribution from the one-loop
chargino contribution in the NMSSM-nuSS. New contribution here means
the difference between the one-loop chargino contribution in the NMSSM-nuSS
and in the NMSSM. It can be obtained by using the mass insertion
method. The Feynman diagram in \figref{fig:OL-AMM-EN} exemplifies the
enhancement mechanism. 
 In the region of large $y_{22}^{\nu}$, the approximate new
 contribution is given by  
\bea 
a_\mu^{\chi^\pm,1l}(\text{new}) &\sim& -\fr{m_l}{8\pi^2}
(y_{22}^{\nu})^2y_\mu  \fr{\mu_{\text{eff}}^2v\Re(s_\beta
  A^\nu_{22}e^{i\varphi_u} - c_\beta \mu_{\text{eff}} ) }{ 
(M_{\ti \nu_+\ti \nu_+})_{22} -(M_{\ti N_+\ti N_+})_{22}}\crn
&& (D_1((M_{\ti \nu_+\ti
  \nu_+})_{22},\mu_{\text{eff}}^2,\mu_{\text{eff}}^2,\mu_{\text{eff}}^2)- 
D_1((M_{\ti N_+\ti
  N_+})_{22},\mu_{\text{eff}}^2,\mu_{\text{eff}}^2,\mu_{\text{eff}}^2)
), \label{eq:approx} 
\eea
where $D_1$ denotes the rank-1 four-point function where all external
momenta are set equal to zero, 
\be
D_1(x,y,y,y) = -\fr{(2x^2+ 5xy-y^2)}{12(x-y)^3y} + \fr{x^2\log(x/y)}{2(x-y)^4}\,,
 \ee and $(M_{\ti \nu_+\ti
  \nu_+})_{22},(M_{\ti N_+\ti N_+})_{22}$ are the second diagonal
components of the sneutrino mass matrix, see
\appen{appen:sneumass}. This contribution is  
proportional to $(y_{22}^{\nu})^2$ in which one factor $y_{22}^{\nu}$ arises
from the coupling $\ti {H_u} \mu \ti N $  and the other comes from 
the mixing between $\ti \nu$ and $\ti N$. 
In the upper left plot \figref{fig:ynu}, we also highlighted the
1$\sigma$ (light gray) and the 2$\sigma$ (dark gray) regions of the difference
between the experimental value and the SM prediction as defined in
\eqref{eq:anomaly}. The points denoted by green triangles are those
points that pass all our constraints. In the lower left plot of \figref{fig:ynu}
we show the relative difference between the muon AMM  in the two
models NMSSM and NMSSM-nuSS, defined as 
\be 
\Delta_{a_\mu} = \bigg|\fr{(a^{x}_\mu(\text{NMSSM-nuSS})- a^{x}_\mu(\text{NMSSM}) )
}{a^{x}_\mu(\text{NMSSM})}\bigg| \,,
\ee
where $x$ can be the chargino one-loop contribution  or the sum of all
contributions. The relative difference is dominated by the
  chargino one-loop contribution and strongly increases with
  $y_{22}^\nu$ from 0 to more than 350\% in the range of the $y_{22}^\nu$ variation.
In the upper right plot of \figref{fig:ynu}  we show the variation of
the loop-corrected Higgs boson mass for the $h_u$-like state at order
$\orderO{\alpha_t \alpha_s +  \alpha_t^2}$ as a function of
$y_{22}^\nu$.  As can be inferred from the plot, in the region
$y_{22}^{\nu}>0.65$ the neutrino/sneutrino sector  
  strongly affects the mass of the $h_u$-like Higgs boson.
 It increases until $y_{22}^{\nu}$ reaches  $1.2$ and then quickly
 decreases. This is due to the interplay between the positive
 contributions from the neutrino one-loop diagrams and the negative 
  contributions from the sneutrino one-loop diagrams. 
  The variation of $y_{22}^{\nu}$ affects both contributions
  simultaneously. At large $y_{22}^{\nu}$ the sneutrino mass becomes
  very small so that its effect gets stronger than the neutrino one
  and it reduces the mass $M_{h_u}$ of the $h_u$-like Higgs boson to a
  very small value.  The relative difference between the $h_u$-like
  Higgs boson mass in the NMSSM-nuSS and the NMSSM as function of
  $y_{22}^{\nu}$ is shown in the lower right plot of
  \figref{fig:ynu}. From small values it increases starting from
    $y_{22}^\nu=0.65$ until it reaches a maximum of 3\% at 1.2 and
    decreases again to small relative differences.
 %
%%%%%%%%%%
\begin{figure}[t]
    \centering
    \subfloat[]{
    \begin{tabular}[b]{r}
        \includegraphics[width=0.495\textwidth]{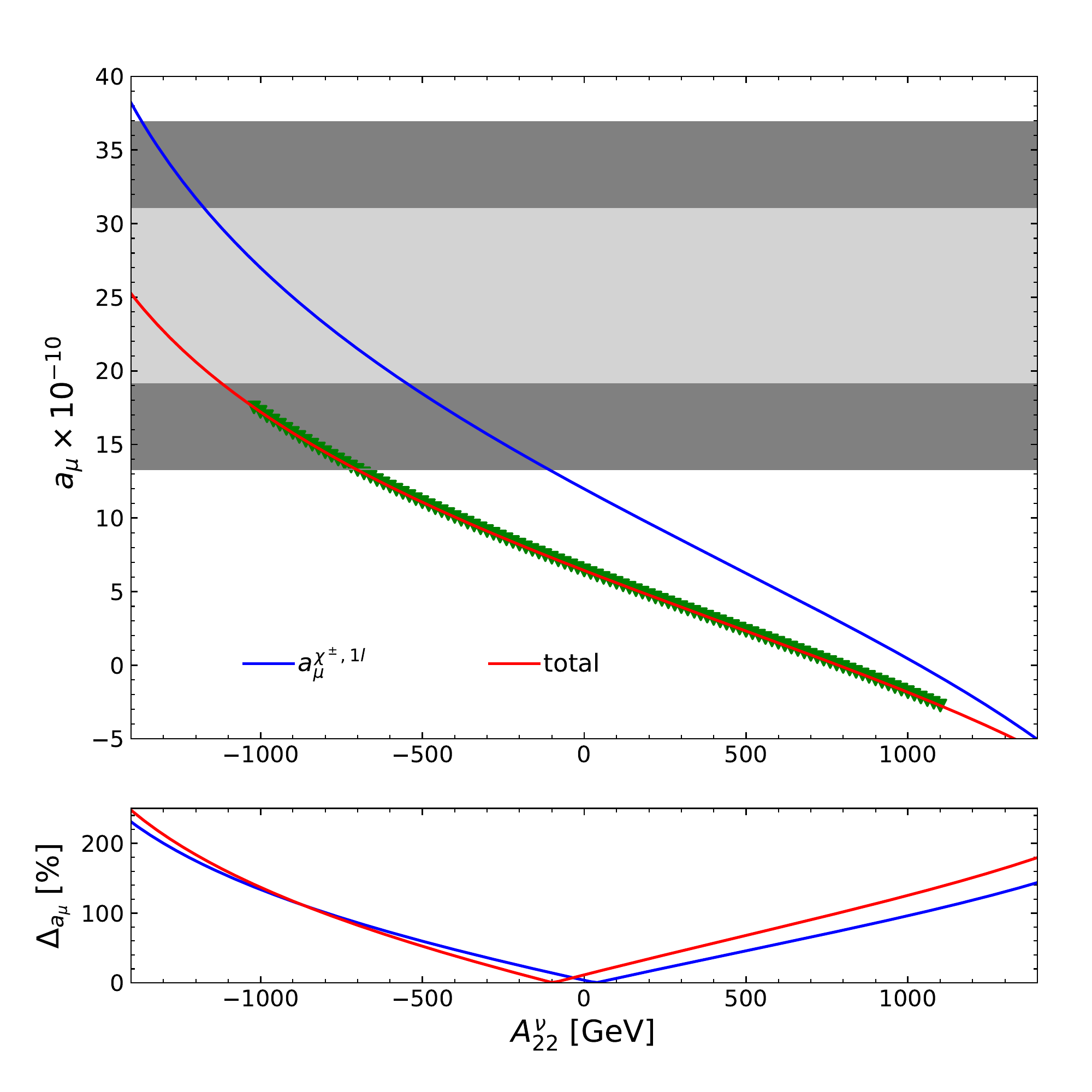}
    \end{tabular}
    }
    \subfloat[]{
    \begin{tabular}[b]{r}
        \includegraphics[width=0.495\textwidth]{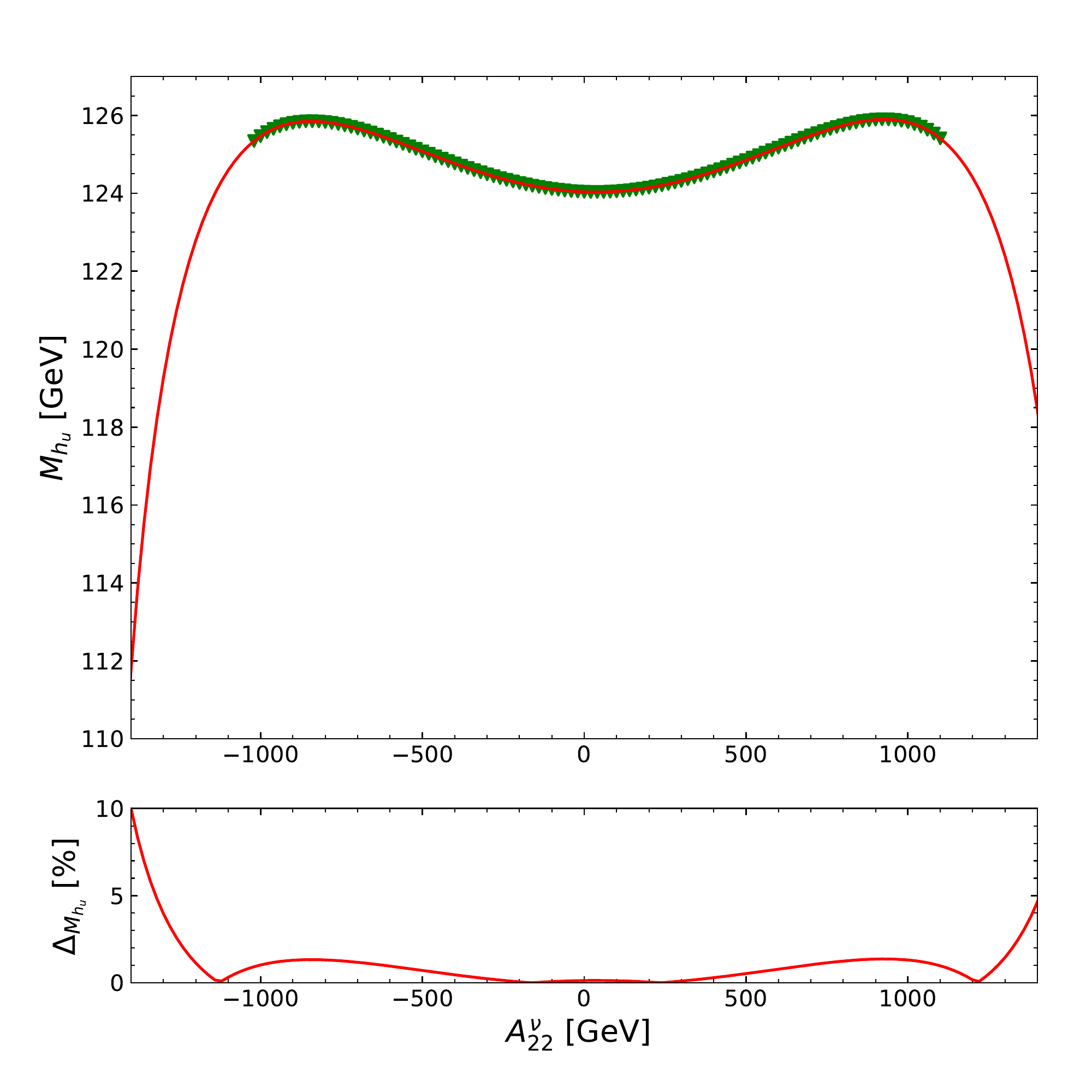}
    \end{tabular}
    }
    \caption{Similar to \figref{fig:ynu} but now $A^\nu_{22}$ is
      varied instead.
  } 
    \label{fig:Anu}
\end{figure}
%%%%%%%%%% 

%%%%%%%%%%
\begin{figure}[t]
    \centering
    \subfloat[]{
    \begin{tabular}[b]{r}
        \includegraphics[width=0.495\textwidth]{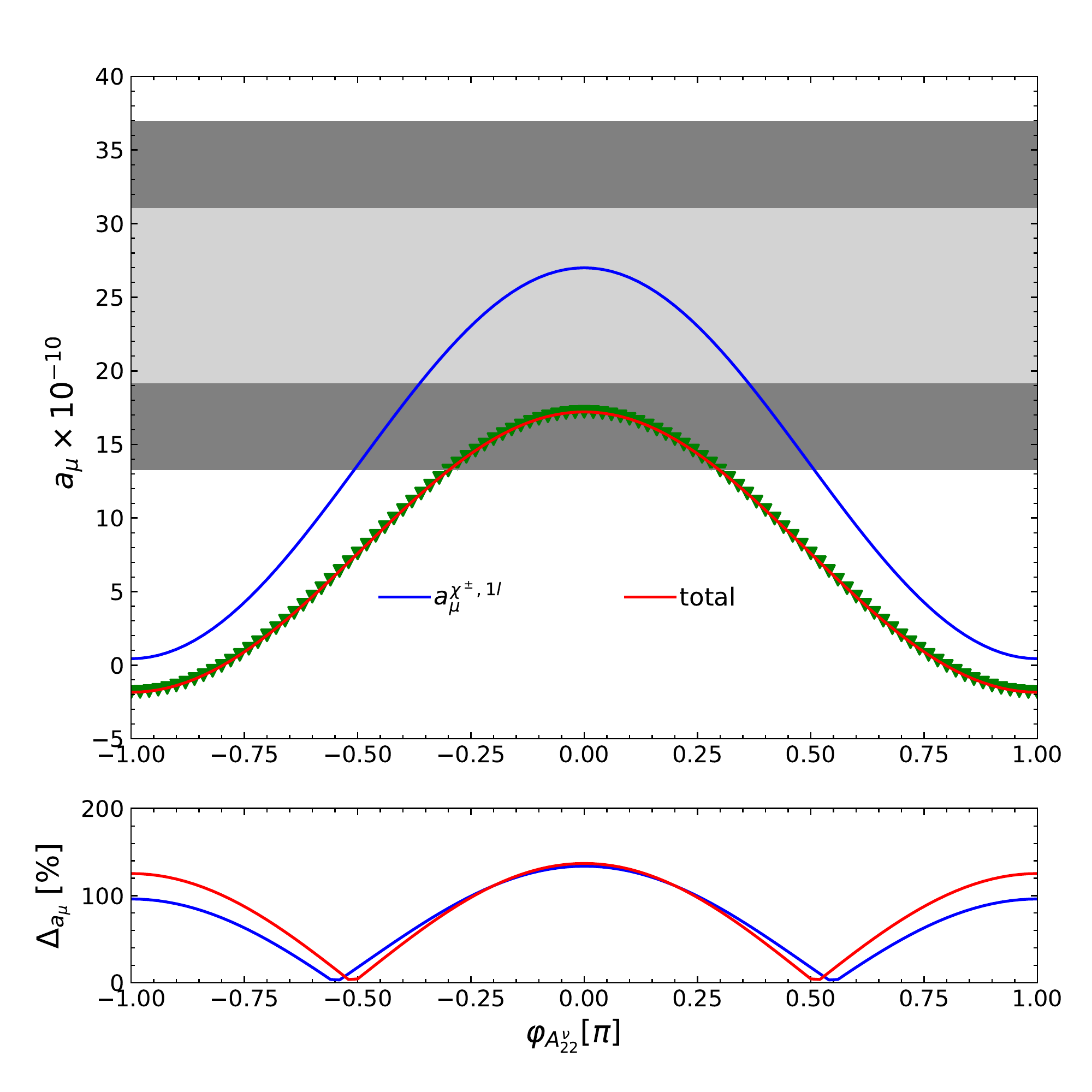}
    \end{tabular}
    }
    \subfloat[]{
    \begin{tabular}[b]{r}
        \includegraphics[width=0.495\textwidth]{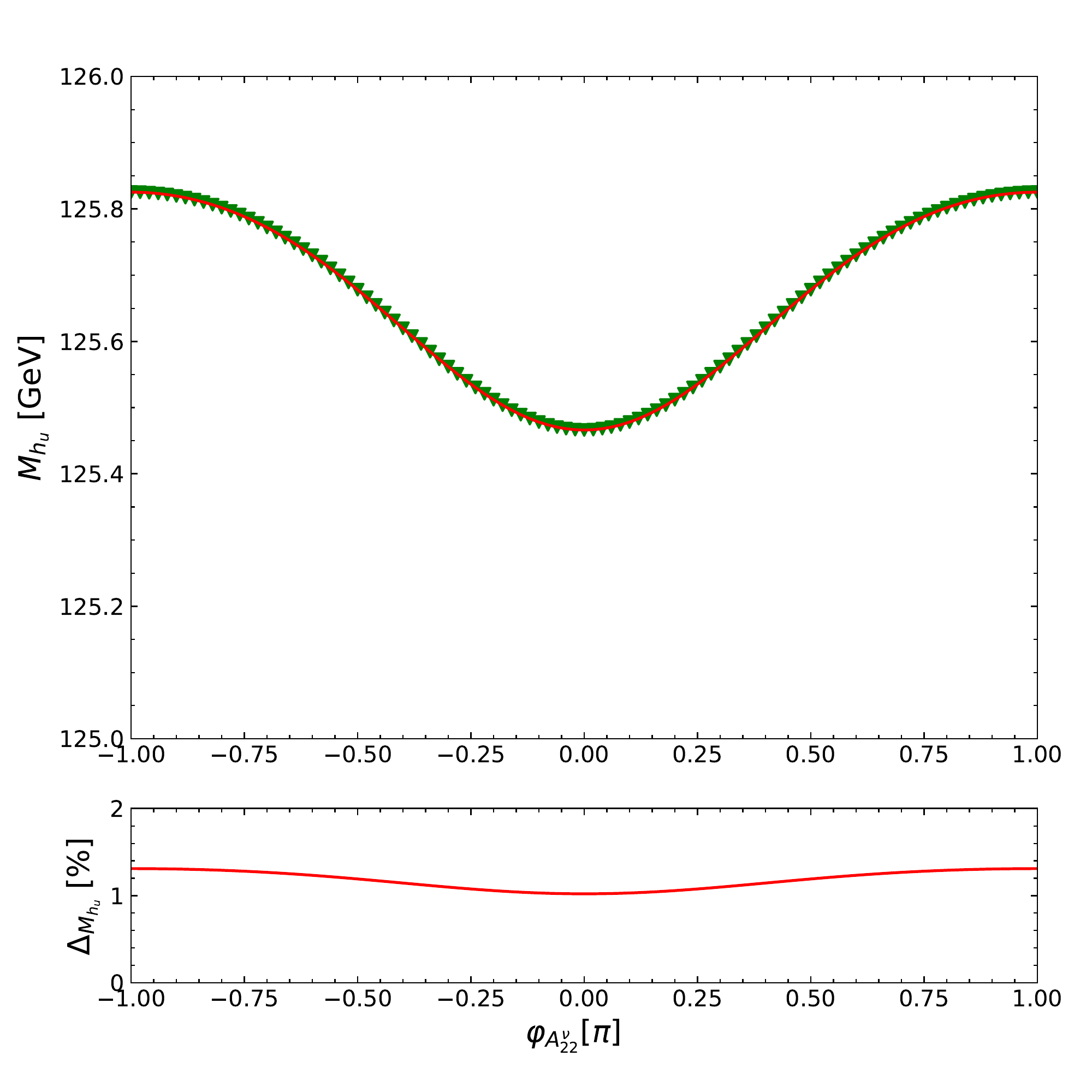}
    \end{tabular}
    }
    \caption{Effect of  the complex phase of $A^\nu_{22}$ on the
      AMM of the muon (left) and the loop-corrected mass of the $h_u$-like
      Higgs boson. The color code is the same as \figref{fig:ynu}.
  } 
    \label{fig:AnuCP}
\end{figure}
%%%%%%%%%% 
 
\paragraph{Dependence on $A_{22}^\nu$:}
 The dependence of the muon AMM  and the loop-corrected
 $h_u$-like Higgs boson mass on the  magnitude of the neutrino
 soft SUSY-breaking trilinear coupling $A_{22}^\nu$ is presented in
 \figref{fig:Anu}. We varied $A_{22}^\nu$ in the range
 $[-1400,1400]\,\gev$. The notation and color code is the same as in
 \figref{fig:ynu}. The nearly linear dependence of the chargino
 one-loop contribution seen in \figref{fig:Anu} (upper left) can be
 explained by using the approximate expression in
 \eqref{eq:approx}. The change of the sign of the new contribution
 around $A_{22}^\nu\sim 0$ can be seen in the lower left plot of
 \figref{fig:Anu}. 
For the explanation of the experimental result for $(g-2)_\mu$ a
negative value of $A_{22}^\nu$ is preferred. This feature
gives also a possibility for the NMSSM-nuSS to explain simutaneously
both the positive discrepancy in $(g-2)_\mu$  and  the negative
discrepancy in $(g-2)_e$ \cite{Aoyama:2017uqe,Hanneke_2008,PhysRevA.83.052122} 
 by choosing a negative value for  $A_{22}^\nu$ and a positive value for $A_{11}^\nu$ as shown 
 in \cite{Cao:2021lmj}. 
For the parameter point {\tt P1}, it is impossible to obtain 
the SUSY contributions for the electron AMM of $-7\times 10^{-13}$  to be close to
the deviation between the experimental measurement and the SM
prediction while it still satisfies other constraints. 
In the right plots of \figref{fig:Anu}, we can see the
dependence of the loop-corrected $h_u$-like Higgs boson mass on
$A_{22}^\nu$ in the upper plot while in the lower plot we see the
relative difference of this mass in the two models with and without
inverse seesaw mechanism. The larger the magnitude of $A_{22}^\nu$
  is, the larger the mixing between $\ti \nu$ and $\ti N$ becomes. This
  leads to the reduction of the mass of the left-handed muon-like 
 sneutrino. As a consequence the sneutrino contributions become
dominant compared to the neutrino contributions.  
   %

%%%%%%%%%%
\begin{figure}[htb!]
    \centering
    \subfloat[]{
    \begin{tabular}[b]{r}
        \includegraphics[width=0.495\textwidth]{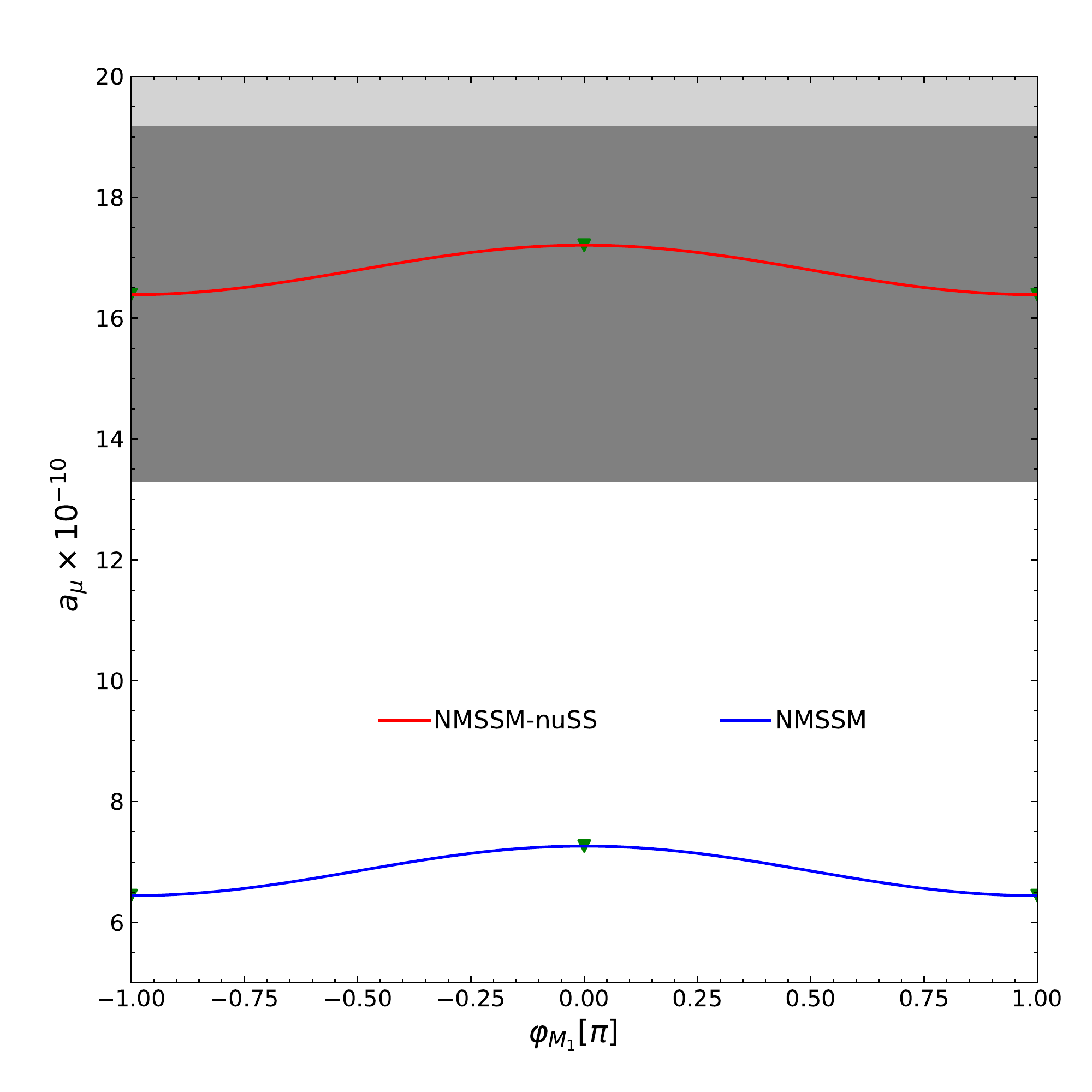}
    \end{tabular}
    }
    \subfloat[]{
    \begin{tabular}[b]{r}
        \includegraphics[width=0.495\textwidth]{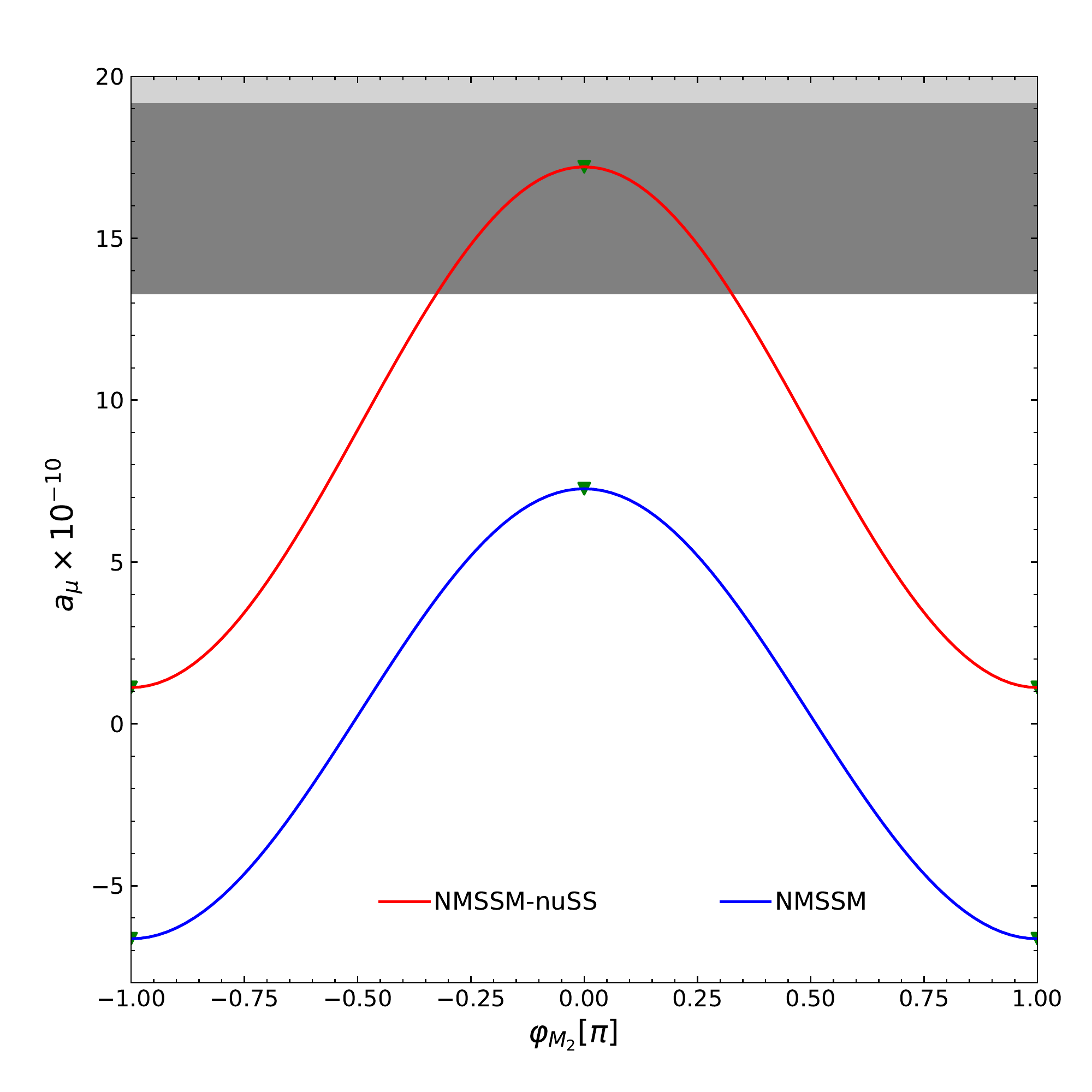}
    \end{tabular}
    }  \\
    \subfloat[]{
    \begin{tabular}[b]{r}
        \includegraphics[width=0.495\textwidth]{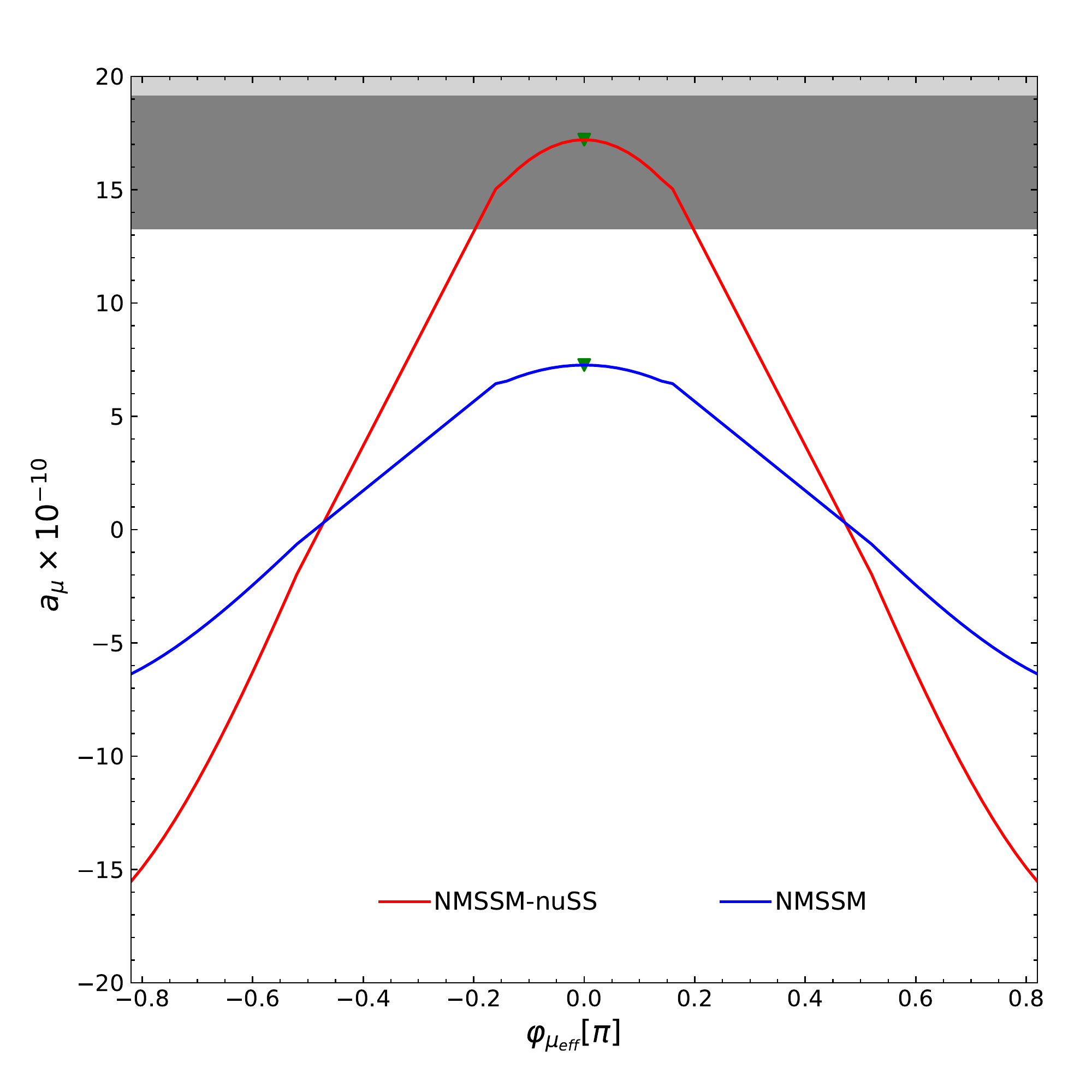}
    \end{tabular}
    } 
    \subfloat[]{
    \begin{tabular}[b]{r}
        \includegraphics[width=0.495\textwidth]{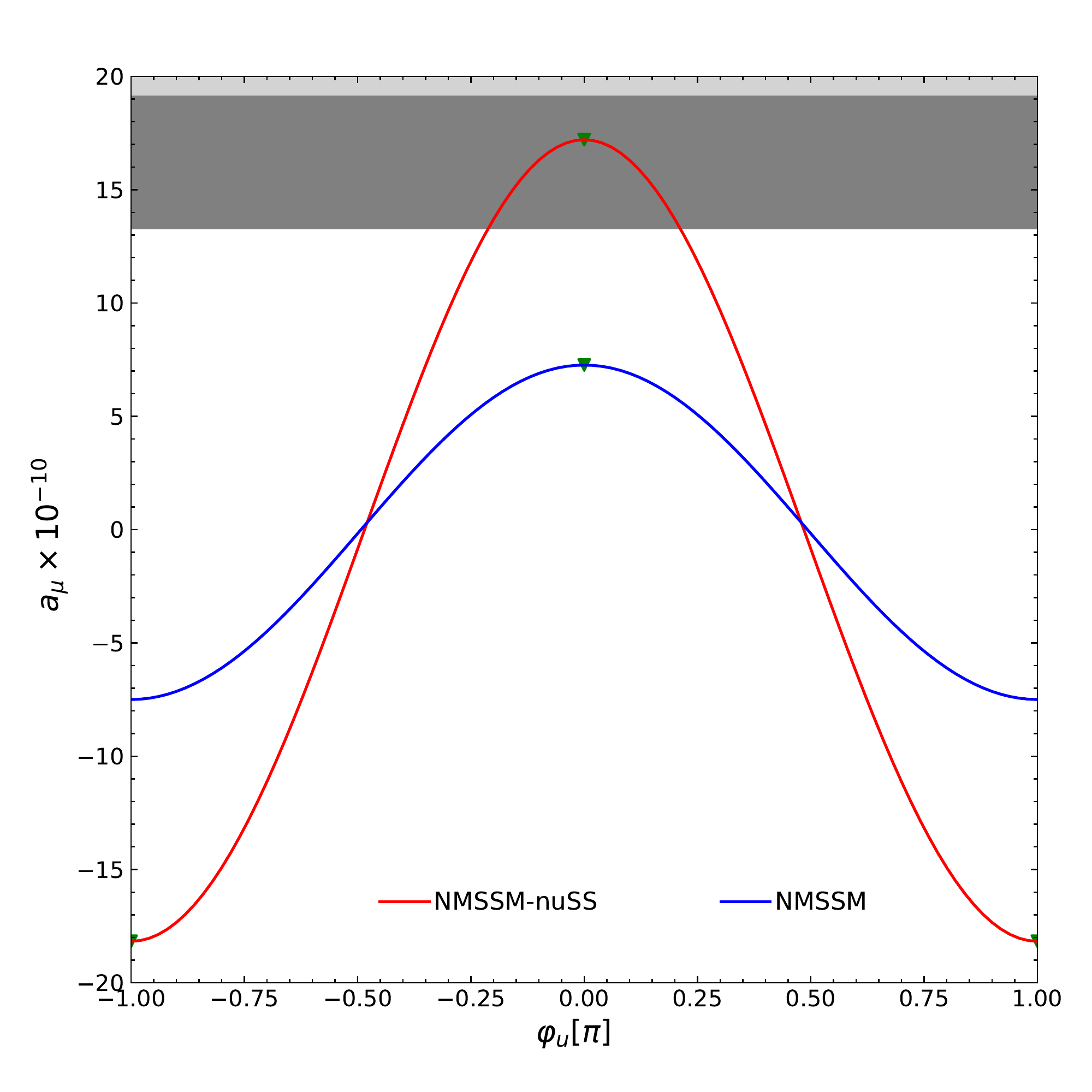}
    \end{tabular}
    }       
    \caption{The AMM of the muon in the NMSSM with (red) and without
      (blue) inverse seesaw mechanism as a function of several
      CP-violating phases: (a) $\varphi_{M_1}$, (b) $\varphi_{M_2}$, (c)
      $\varphi_{\mu_{\text{eff}}}$, (d) $\varphi_u$. 
  } 
    \label{fig:CPphases}
\end{figure}

\paragraph{Influence of the CP-violating phases:}
We now discuss the influence of the CP-violating phases on the muon AMM 
 and the loop-corrected $h_u$-like Higgs boson mass. In
\figref{fig:AnuCP}, we vary the complex phase of  $A^\nu_{22}$ in the
range $[-\pi, \pi]$. The SUSY contributions to $a_\mu$ change its
value from $-2\times 10^{-10}$ at $\varphi_{A^\nu_{22}}=-\pi$ to reach a
maximum of $17.2\times 10^{-10}$ at $\varphi_{A^\nu_{22}}=0$ and then reduce
it back to $-2\times 10^{-10}$ at $\varphi_{A^\nu_{22}}=\pi$.  While the complex
 phase of  $A^\nu_{22}$ strongly affects the $(g-2)_\mu$, its effect
 on the mass of the $h_u$-like Higgs boson is rather mild as can seen
 in the right plots of \figref{fig:AnuCP}. 
We further present in \figref{fig:CPphases}
the influence of several complex phases, namely
$\varphi_{M_1},\varphi_{M_2},\varphi_{\mu_\text{eff}},\varphi_u$, on
the SUSY contributions to $a_\mu$ in both models, 
the NMSSM with and without inverse seesaw mechanism. In all these
plots, the NMSSM-nuSS results are 
plotted in red while the blue lines show the results in the NMSSM
without inverse seesaw mechanism.  
The complex phase of $M_1$ enters only the neutralino
contribution. Figure~\ref{fig:CPphases} (a) shows a mild dependence of
$(g-2)_\mu$ on this phase for this particular point where the
neutralino contribution is always negative and about four times
smaller than the dominant chargino contribution. The complex phase of
$M_2$ enters not only the neutralino contribution but also the
chargino one. In \figref{fig:CPphases} (b)  we can see a similar
dependence of $(g-2)_\mu$ on this phase in both models. The two remaining phases $\varphi_{\mu_\text{eff}}$ and $\varphi_{u}$ have a stronger influence on
  $(g-2)_\mu$ in the NMSSM-nuSS than in the NMSSM without ISS as shown
  in \figref{fig:CPphases} (c) and \figref{fig:CPphases} (d). This is
  due to these two phases entering the new contribution in the
  NMSSM-nuSS, see \eqref{eq:approx}. Note that  in
  \figref{fig:CPphases} (c) the range of $\varphi_{\mu_\text{eff}}$ is
  $[-0.8\pi,08\pi]$ since outside this range the mass of the
  $h_u$-like Higgs boson turns out to be negative.  For
  illustrative purpose we show partly the light gray and dark gray
  regions representing the 1$\sigma$ and 2$\sigma$ deviations between
  the experimental measurement and the SM prediction for
  $(g-2)_\mu$. In these plots,  except for the points where the phases
  $\varphi_{M_1},\phi_{M_2},\varphi_{\mu_\text{eff}},\varphi_u$ are close to zero or
$\pm\pi$, all other points are ruled out because of the constraints on
the electric dipole moments of the electron.

\paragraph{Effects on the electron EDM:}
We now investigate the effect of the new complex phase in the
  (s)neutrino sectors on the electron EDM. In {\tt NMSSMCALC}, the electron,
  neutron, Thallium and Mercury EDMs have been implemented as
  described in \cite{King:2015oxa}. We follow the conventions in {\tt NMSSMCALC}
that all EDMs are normalized to their corresponding experimental upper
bounds. A thorough investigation of the complex phases in the NMSSM on
the electron, neutron, Thallium and Mercury EDMs in
\cite{King:2015oxa} has shown that the complex phases of the
electroweak sector such as $\varphi_{M_1},\varphi_{M_2},$
$\varphi_{1}\equiv \varphi_\lambda+\varphi_s+\varphi_u, \varphi_2
\equiv \varphi_\kappa + 3 \varphi_s$ have the 
strongest effects on the electron EDM which also enters the Thallium and
Mercury EDMs. These phases contribute to the electron EDM through the
one-loop neutralino and chargino contributions as can be inferred from
the second terms in \eqref{eq:olneutralino} and \eqref{eq:olchargino}. 
Apart from the NMSSM-like phase $\varphi_2$, the stringent limit
on the electron EDM has ruled out almost any non-vanishing value of 
these complex phases as we also observe in this study (see 
previous paragraph).
In the NMSSM-nuSS, there are new complex phases from the neutrino sector, 
$\delta_{CP}, \varphi_{y^\nu_{11}}, \varphi_{\mu^X_{11}}$,
$\varphi_{\lambda^X_{11}}$, and from the sneutrino sector
$\varphi_{A^\nu_{11}}, \varphi_{A^X_{11}},
\varphi_{B^{\mu_X}_{11}}$. However, only the complex phase
$\varphi_{A^\nu_{11}}$ gives a significant contribution to the EDMs, all
 other remaining phases have a negligible effect. The complex phase
 $\varphi_{A^\nu_{11}}$ appears in the one-loop chargino contribution,
 see the second term in \eqref{eq:olchargino}. This provides a
 possibility for the reduction of the imarginary part of the coupling
 $g^R_{l\ti \chi^-_j \tilde{\nu}_i}$, see \eqref{eq:charginoCL}, and
 may thereby lead to the cancellation between different contributions
 to the electron EDM. Such a cancellation can never happen in the
 NMSSM without ISS. 

\begin{figure}[t]
    \centering
    \subfloat[]{
    \begin{tabular}[b]{r}
        \includegraphics[width=0.495\textwidth]{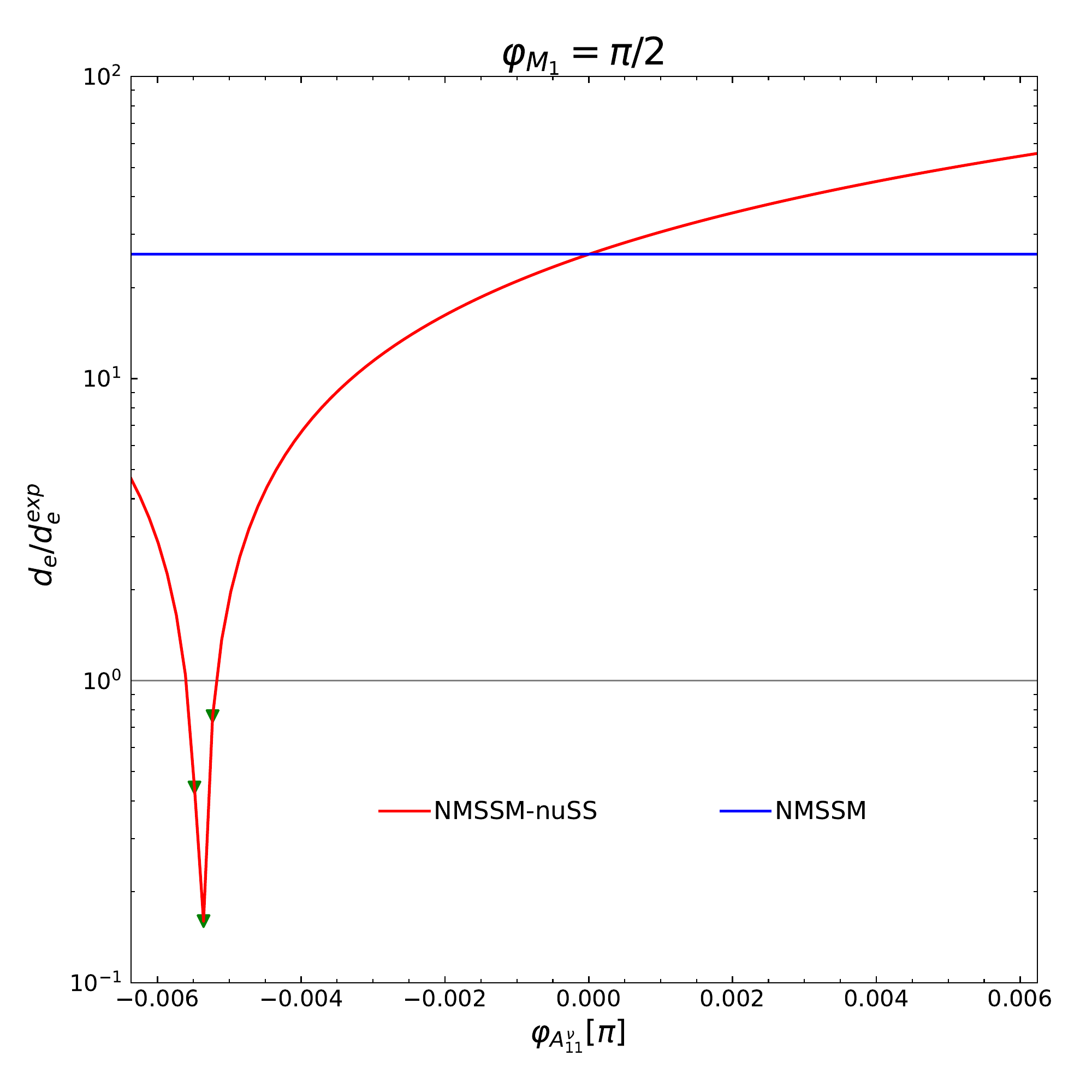}
    \end{tabular}
    }
    \subfloat[]{
    \begin{tabular}[b]{r}
        \includegraphics[width=0.495\textwidth]{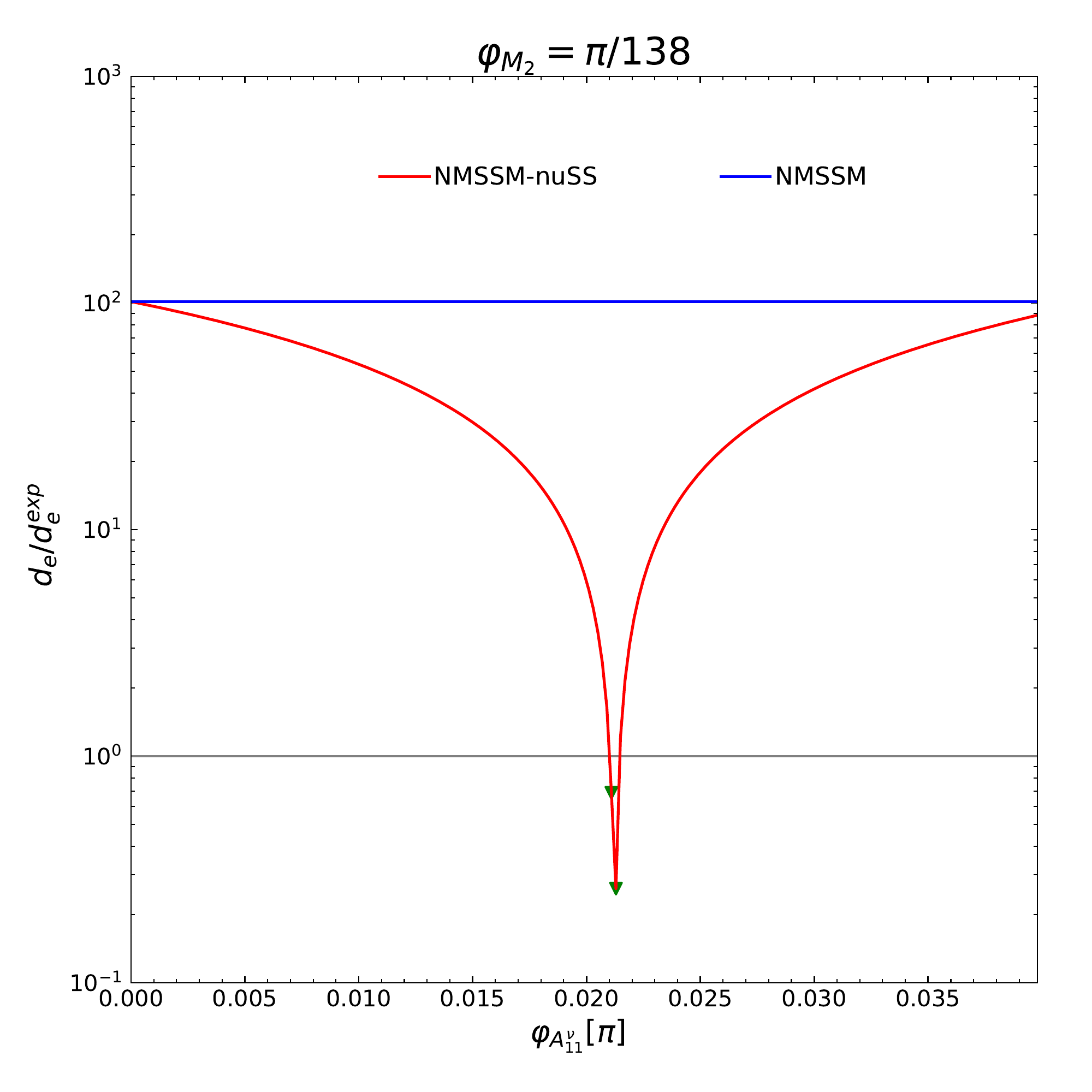}
    \end{tabular}
    }       
    \caption{The EDM of the electron in the NMSSM with (red) and without
      (blue) inverse seesaw mechanism as a function of the complex phase
      of $A^\nu_{11}$ where (a) $\varphi_{M_1}=\pi/2$, (b) $\varphi_{M_2}=\pi/138$. 
  } 
    \label{fig:CPphasesEDM}
\end{figure}

To illustrate this, we show in
 \figref{fig:CPphasesEDM} the dependence of the electron EDM,
 normalized to the experimental upper bound, on the complex phase of
 $A^\nu_{11}$ using the parameter point {\tt P1}. In the left plot, we
 have set   $\varphi_{M_1}=\pi/2$ while in the right plot we have set
 $\varphi_{M_2}=\pi/138$. All other phases are equal to zero. These
 values of $\varphi_{M_1}$ and $\varphi_{M_2}$ correspond to  the largest
 possible induced EDM of the  
 neutron  
from the respective phases that still remains below the experimental upper bound.
  The red (blue) lines show the electron EDMs in the NMSSM with
  (without) ISS. As can be inferred from the plots, in the NMSSM
  without ISS the electron EDM
  is about  $25$ times larger than its experimental upper bound  for
  $\varphi_{M_1}=\pi/2$ and about 102 times larger for  
  $\varphi_{M_2}=\pi/138$.  In the NMSSM with ISS a cancellation of
  all  contributions to the electron EDM takes
  place at $\varphi_{A_{11}^\nu}= -0.0055\pi$ for
  $\varphi_{M_1}=\pi/2$ and at $\varphi_{A_{11}^\nu}= 0.022\pi$ for
  $\varphi_{M_2}=\pi/138$ so that the electron EDM is pushed below its
  experimental upper bound. Note that all the points with the electron
  EDM being less than one satisfy all our constraints mentioned in
  this paper. A similar cancellation can also happen for the phases
  $\varphi_{1}$ and $\varphi_{2}$.

%%%%%%%%%%   
%%%%%%%%%%%%%%%%%%%%%%%%%%%%%%%%%%%%%%%%%%%%%%%%%%%%%%%%%%%%%%
\section{Conclusions}
\label{sec:conclusions}
%%%%%%%%%%%%%%%%%%%%%%%%%%%%%%%%%%%%%%%%%%%%%%%%%%%%%%%%%%%%%%
In this  paper we have computed the full one-loop SUSY contributions
and the two-loop Barr-Zee-type diagrams with $h\gamma\gamma$ effective 
couplings to the AMM and EDM of charged leptons in two models, the NMSSM with and without inverse
seesaw mechanism including CP-violating phases. 
We presented the analytic expressions and implemented them in the two
Fortran codes {\tt NMSSMCALC} and {\tt NMSSMCALC-nuSS}, 
which compute the Higgs boson masses and mixings, together with the Higgs boson
decay branching ratios taking into account the most up-to-date
higher-order corrections. 
Using a typical parameter point with an intermediate
value of $\tan\beta$ and large charged Higgs mass, we have
investigated in the NMSSM with inverse 
seesaw mechanism the effect of the (s)neutrino sector on the muon 
AMM in comparison with its effect
on the SM-like Higgs-boson mass. We see a large positive contribution
to the AMM from the mixing between
the left-handed muon-type sneutrino and the right-handed one (denoted
as $\ti N$ in the previous sections) provided that the muon-type
neutrino Yukawa coupling is of order ${\cal O}(1)$, that the muon-type neutrino trilinear
 coupling is negative and that the left- and right-handed
 muon-type sneutrino masses are small. Too light sneutrino
   masses, however, give a large negative correction to the Higgs
   boson masses. In order to compensate this negative effect one should also require 
   light sterile neutrino masses. Therefore, there is a strong correlation between the
 effects of the (s)neutrino sector on the two observables. 
 We have also found a strong effect of the CP-violating phases on the
 AMM of the muon in the two models. 
 
 For the electron EDM we found that the complex phase of the sneutrino sector
  $\varphi_{A^\nu_{11}}$ gives a significant contribution at one-loop level. This provides a 
  possibility for the cancellation of the different contributions to the electron EDM
so that it remains below the experimental upper bound. While most of the
non-vanishing complex phases of  
the electroweak sector of the NMSSM have been ruled out by the constraint on  the 
electron EDM, in the NMSSM-nuSS one can remain in the region of
validity with an appropriately chosen value of $\varphi_{A^\nu_{11}}$. 

Finally, the calculations presented in this paper have been implemented in the programs 
{\tt NMSSMCALC} and {\tt NMSSMCALC-nuSS} which are publicly available.  

%%%%%%%%%%%%%%%%%%%%%%%%%%%%%%%%%%%%%%%%%%%%%%%%%%%%%%%%%%%%%%
\section*{Acknowledgements}
%%%%%%%%%%%%%%%%%%%%%%%%%%%%%%%%%%%%%%%%%%%%%%%%%%%%%%%%%%%%%%
T.N.D and D.N.L are funded by the Vietnam National Foundation for Science
and Technology Development (NAFOSTED) under grant number 103.01-2020.17. 
The research of MM was supported by the Deutsche
Forschungsgemeinschaft (DFG, German Research Foundation) under grant
396021762 - TRR 257.

%%%%%%%%%%%%%%%%%%%%%%%%%%%%%%%%%%%%%%%%%%%%%%%%%%%%%%%%%%%%%%
\begin{appendix}
%-------------------------------------------------
\section{The Sneutrino Mass Matrix} \label{appen:sneumass}
Here we provide the mass matrix of the sneutrinos. Each entry is a 
$3\times 3$ matrix in the flavor space.  
\begin{align}
(M_\tn)_{\tn_+ \tn_+} &= \half I_3 M_z^2 \cos 2\beta + \half \braket{\ti m_L^2 + \ti m_L^{2T}} + \frac{1}{2} v_u^2 \Re \braket{y_\nu y_\nu^\dagger}\\
(M_\tn)_{\tn_+ \tN_+} &= \frac{1}{\sqrt{2}} v_u \Re\braket{e^{i\vp_u} y_\nu A_\nu} - \frac{1}{2} v_d v_s \Re\braket{e^{i\vp_s} \lambda y_\nu^* }\\
(M_\tn)_{\tn_+ \tX_+} &= \frac{1}{\sqrt{2}} v_u \Re\braket{ e^{i\vp_u} y_\nu \mu_X^* }\\
(M_\tn)_{\tn_+ \tn_-} &= \frac{i}{2} \braket{\ti m_L^2 - \ti m_L^{2T}} + \frac{1}{2} v_u^2 \Im \braket{y_\nu y_\nu^\dagger}\\
(M_\tn)_{\tn_+ \tN_-} &= \frac{1}{\sqrt{2}} v_u \Im\braket{e^{i\vp_u} y_\nu A_\nu} - \frac{1}{2} v_d v_s \Im\braket{e^{i\vp_s} \lambda y_\nu^* }\\
(M_\tn)_{\tn_+ \tX_-} &= \frac{1}{\sqrt{2}} v_u \Im\braket{ e^{i\vp_u} y_\nu \mu_X^* }\\
(M_\tn)_{\tN_+ \tN_+} &= \half\braket{\ti m^2_N + \ti m^{2T}_N} + \Re\braket{\mu_X \mu_X^\dagger} + \frac{1}{2} v_u^2 \Re \braket{y_\nu^T y_\nu^*}\\
(M_\tn)_{\tN_+ \tX_+} &= \Re\braket{\mu_X B_{\mu_X}} + \frac{1}{\sqrt{2}} v_s \Re\left[ e^{-i\vp_s}\mu_X\braket{\lambda_X^\dagger + \lambda_X^*} \right]\\
(M_\tn)_{\tN_+ \tn_-} &= -\frac{1}{\sqrt{2}} v_u \Im\braket{e^{i\vp_u}A_\nu^T y_\nu^T} - \frac{1}{2} v_d v_s \Im\braket{e^{i\vp_s} \lambda y_\nu^\dagger }\\
(M_\tn)_{\tN_+ \tN_-} &= \frac{i}{2}\braket{\ti m^2_N - \ti m^{2T}_N} - \Im\braket{\mu_X \mu_X^\dagger} - \frac{1}{2} v_u^2 \Im \braket{y_\nu^T y_\nu^*}\\
(M_\tn)_{\tN_+ \tX_-} &= -\Im\braket{\mu_X B_{\mu_X}} + \frac{1}{\sqrt{2}} v_s \Im\left[ e^{-i\vp_s}\mu_X\left(\lambda_X^\dagger + \lambda_X^*\right) \right]\\
(M_\tn)_{\tX_+ \tX_+} &= \half \braket{\ti m_X^2 + \ti m_X^{2T}}+ \Re\braket{\mu_X^T \mu_X^*} + \half \Re\left[ \left(e^{2i\vp_s}v_s^2 \kappa - e^{i\vp_u}v_d v_u \lambda\right) \left(\lambda_X^* + \lambda_X^\dagger\right)\right] \notag\\
&\quad+ \half v_s^2 \Re\left[ \left(\lambda_X + \lambda_X^T \right)\left(\lambda_X^\dagger + \lambda_X^*\right) \right] + \frac{1}{\sqrt{2}} v_s \Re \left[e^{i\vp_s}\left(\lambda_X A_X + A_X^T\lambda_X^T\right)\right] \\
(M_\tn)_{\tX_+ \tn_-} &= -\frac{1}{\sqrt{2}} v_u \Im\left( e^{i\vp_u} \mu_X^\dagger y_\nu^T \right)\\
(M_\tn)_{\tX_+ \tN_-} &= \Im\left(B_{\mu_X}^T \mu_X^T\right) + \frac{1}{\sqrt{2}} v_s \Im\left[ e^{-i\vp_s} \left(\lambda_X^\dagger + \lambda_X^*\right) \mu^T_X \right]\\
(M_\tn)_{\tX_+ \tX_-} &= \frac{i}{2} \left(\ti m_X^2 - \ti m_X^{2T}\right) + \Im\left(\mu_X^T \mu_X^*\right) + \half \Im\left[ \left(e^{2i\vp_s}v_s^2 \kappa - e^{i\vp_u}v_d v_u \lambda\right) \left(\lambda_X^* + \lambda_X^\dagger\right)\right] \notag\\
&\quad+ \half v_s^2 \Im\left[ \left(\lambda_X + \lambda_X^T \right)\left(\lambda_X^\dagger + \lambda_X^*\right) \right] - \frac{1}{\sqrt{2}} v_s \Im \left[e^{i\vp_s}\left(\lambda_X A_X + A_X^T\lambda_X^T\right)\right] \\
(M_\tn)_{\tn_- \tn_-} &= \half I_3 M_z^2 \cos 2\beta + \half \left(\ti m_L^2 + \ti m_L^{2T}\right) + \frac{1}{2} v_u^2 \Re \left(y_\nu y_\nu^\dagger\right)\\
(M_\tn)_{\tn_- \tN_-} &= \frac{1}{\sqrt{2}} v_u \Re\left(e^{i\vp_u} y_\nu A_\nu\right) - \frac{1}{2} v_d v_s \Re\left(e^{i\vp_s} \lambda y_\nu^* \right)\\
(M_\tn)_{\tn_- \tX_-} &= \frac{1}{\sqrt{2}} v_u \Re\left( e^{i\vp_u} y_\nu \mu_X^* \right)\\	
(M_\tn)_{\tN_- \tN_-} &= \half\left(\ti m^2_N + \ti m^{2T}_N \right) + \Re\left(\mu_X \mu_X^\dagger\right) + \frac{1}{2} v_u^2 \Re \left(y_\nu^T y_\nu^*\right)
\end{align}
\begin{align}
(M_\tn)_{\tN_- \tX_-} &= \Re\left(\mu_X B_{\mu_X}\right) - \frac{1}{\sqrt{2}} v_s \Re\left[ e^{-i\vp_s}\mu_X\left(\lambda_X^\dagger + \lambda_X^*\right) \right]\\
(M_\tn)_{\tX_- \tX_-} &= \half \left(\ti m_X^2 + \ti m_X^{2T}\right) + \Re\left(\mu_X^T \mu_X^* \right) - \half \Re\left[ \left(e^{2i\vp_s}v_s^2 \kappa - e^{i\vp_u}v_d v_u \lambda\right) \left(\lambda_X^* + \lambda_X^\dagger\right)\right] \notag\\
&\quad+ \half v_s^2 \Re\left[ \left(\lambda_X + \lambda_X^T \right)\left(\lambda_X^\dagger + \lambda_X^*\right) \right] - \frac{1}{\sqrt{2}} v_s \Re \left[e^{i\vp_s}\left(\lambda_X A_X + A_X^T\lambda_X^T\right)\right].
\end{align}

\end{appendix}

%%%%%%%%%%%%%%%%%%%%%%%%%%%%%%%%%%%%%%%%%%%%%%%%%%%%%%%%%%%%%%

\providecommand{\href}[2]{#2}\begingroup\raggedright\endgroup

\begin{thebibliography}{10}

\bibitem{Muong-2:2021ojo}
{\scshape Muon g-2} collaboration, B.~Abi et~al., \emph{{Measurement of the
  Positive Muon Anomalous Magnetic Moment to 0.46 ppm}},
  \href{http://dx.doi.org/10.1103/PhysRevLett.126.141801}{\emph{Phys. Rev.
  Lett.} \textbf{ 126} (2021) 141801},
  [\href{https://arxiv.org/abs/2104.03281}{{\texttt 2104.03281}}].

\bibitem{Muong-2:2006rrc}
{\scshape Muon g-2} collaboration, G.~W. Bennett et~al., \emph{{Final Report of
  the Muon E821 Anomalous Magnetic Moment Measurement at BNL}},
  \href{http://dx.doi.org/10.1103/PhysRevD.73.072003}{\emph{Phys. Rev. D}
  \textbf{ 73} (2006) 072003},
  [\href{https://arxiv.org/abs/hep-ex/0602035}{{\texttt hep-ex/0602035}}].

\bibitem{Aoyama:2020ynm}
T.~Aoyama et~al., \emph{{The anomalous magnetic moment of the muon in the
  Standard Model}},
  \href{http://dx.doi.org/10.1016/j.physrep.2020.07.006}{\emph{Phys. Rept.}
  \textbf{ 887} (2020) 1--166},
  [\href{https://arxiv.org/abs/2006.04822}{{\texttt 2006.04822}}].

\bibitem{Aoyama:2012wk}
T.~Aoyama, M.~Hayakawa, T.~Kinoshita and M.~Nio, \emph{{Complete Tenth-Order
  QED Contribution to the Muon g-2}},
  \href{http://dx.doi.org/10.1103/PhysRevLett.109.111808}{\emph{Phys. Rev.
  Lett.} \textbf{ 109} (2012) 111808},
  [\href{https://arxiv.org/abs/1205.5370}{{\texttt 1205.5370}}].

\bibitem{Chao:2021tvp}
E.-H. Chao, R.~J. Hudspith, A.~G\'erardin, J.~R. Green, H.~B. Meyer and
  K.~Ottnad, \emph{{Hadronic light-by-light contribution to $(g-2)_\mu $ from
  lattice QCD: a complete calculation}},
  \href{http://dx.doi.org/10.1140/epjc/s10052-021-09455-4}{\emph{Eur. Phys. J.
  C} \textbf{ 81} (2021) 651},
  [\href{https://arxiv.org/abs/2104.02632}{{\texttt 2104.02632}}].

\bibitem{Czarnecki:2001pv}
A.~Czarnecki and W.~J. Marciano, \emph{{The Muon anomalous magnetic moment: A
  Harbinger for 'new physics'}},
  \href{http://dx.doi.org/10.1103/PhysRevD.64.013014}{\emph{Phys. Rev. D}
  \textbf{ 64} (2001) 013014},
  [\href{https://arxiv.org/abs/hep-ph/0102122}{{\texttt hep-ph/0102122}}].

\bibitem{Fayet:1974pd}
P.~Fayet, \emph{{Supergauge Invariant Extension of the Higgs Mechanism and a
  Model for the electron and Its Neutrino}},
  \href{http://dx.doi.org/10.1016/0550-3213(75)90636-7}{\emph{Nucl. Phys. B}
  \textbf{ 90} (1975) 104--124}.

\bibitem{Barbieri:1982eh}
R.~Barbieri, S.~Ferrara and C.~A. Savoy, \emph{{Gauge Models with Spontaneously
  Broken Local Supersymmetry}},
  \href{http://dx.doi.org/10.1016/0370-2693(82)90685-2}{\emph{Phys.Lett.}
  \textbf{ B119} (1982) 343}.

\bibitem{Dine:1981rt}
M.~Dine, W.~Fischler and M.~Srednicki, \emph{{A Simple Solution to the Strong
  CP Problem with a Harmless Axion}},
  \href{http://dx.doi.org/10.1016/0370-2693(81)90590-6}{\emph{Phys.Lett.}
  \textbf{ B104} (1981) 199}.

\bibitem{Nilles:1982dy}
H.~P. Nilles, M.~Srednicki and D.~Wyler, \emph{{Weak Interaction Breakdown
  Induced by Supergravity}},
  \href{http://dx.doi.org/10.1016/0370-2693(83)90460-4}{\emph{Phys.Lett.}
  \textbf{ B120} (1983) 346}.

\bibitem{Frere:1983ag}
J.~Frere, D.~Jones and S.~Raby, \emph{{Fermion Masses and Induction of the Weak
  Scale by Supergravity}},
  \href{http://dx.doi.org/10.1016/0550-3213(83)90606-5}{\emph{Nucl.Phys.}
  \textbf{ B222} (1983) 11}.

\bibitem{Derendinger:1983bz}
J.~Derendinger and C.~A. Savoy, \emph{{Quantum Effects and SU(2) x U(1)
  Breaking in Supergravity Gauge Theories}},
  \href{http://dx.doi.org/10.1016/0550-3213(84)90162-7}{\emph{Nucl.Phys.}
  \textbf{ B237} (1984) 307}.

\bibitem{Ellis:1988er}
J.~R. Ellis, J.~Gunion, H.~E. Haber, L.~Roszkowski and F.~Zwirner, \emph{{Higgs
  Bosons in a Nonminimal Supersymmetric Model}},
  \href{http://dx.doi.org/10.1103/PhysRevD.39.844}{\emph{Phys. Rev. D} \textbf{
  39} (1989) 844}.

\bibitem{Drees:1988fc}
M.~Drees, \emph{{Supersymmetric Models with Extended Higgs Sector}},
  \href{http://dx.doi.org/10.1142/S0217751X89001448}{\emph{Int. J. Mod. Phys.
  A} \textbf{ 4} (1989) 3635}.

\bibitem{Ellwanger:1993xa}
U.~Ellwanger, M.~Rausch~de Traubenberg and C.~A. Savoy, \emph{{Particle
  spectrum in supersymmetric models with a gauge singlet}},
  \href{http://dx.doi.org/10.1016/0370-2693(93)91621-S}{\emph{Phys. Lett. B}
  \textbf{ 315} (1993) 331--337},
  [\href{https://arxiv.org/abs/hep-ph/9307322}{{\texttt hep-ph/9307322}}].

\bibitem{Ellwanger:1995ru}
U.~Ellwanger, M.~Rausch~de Traubenberg and C.~A. Savoy, \emph{{Higgs
  phenomenology of the supersymmetric model with a gauge singlet}},
  \href{http://dx.doi.org/10.1007/BF01553993}{\emph{Z. Phys. C} \textbf{ 67}
  (1995) 665--670}, [\href{https://arxiv.org/abs/hep-ph/9502206}{{\texttt
  hep-ph/9502206}}].

\bibitem{Ellwanger:1996gw}
U.~Ellwanger, M.~Rausch~de Traubenberg and C.~A. Savoy, \emph{{Phenomenology of
  supersymmetric models with a singlet}},
  \href{http://dx.doi.org/10.1016/S0550-3213(97)00128-4}{\emph{Nucl.Phys.}
  \textbf{ B492} (1997) 21--50},
  [\href{https://arxiv.org/abs/hep-ph/9611251}{{\texttt hep-ph/9611251}}].

\bibitem{Elliott:1994ht}
T.~Elliott, S.~King and P.~White, \emph{{Unification constraints in the
  next-to-minimal supersymmetric standard model}},
  \href{http://dx.doi.org/10.1016/0370-2693(95)00381-T}{\emph{Phys.Lett.}
  \textbf{ B351} (1995) 213--219},
  [\href{https://arxiv.org/abs/hep-ph/9406303}{{\texttt hep-ph/9406303}}].

\bibitem{King:1995vk}
S.~King and P.~White, \emph{{Resolving the constrained minimal and
  next-to-minimal supersymmetric standard models}},
  \href{http://dx.doi.org/10.1103/PhysRevD.52.4183}{\emph{Phys.Rev.} \textbf{
  D52} (1995) 4183--4216},
  [\href{https://arxiv.org/abs/hep-ph/9505326}{{\texttt hep-ph/9505326}}].

\bibitem{Franke:1995tc}
F.~Franke and H.~Fraas, \emph{{Neutralinos and Higgs bosons in the
  next-to-minimal supersymmetric standard model}},
  \href{http://dx.doi.org/10.1142/S0217751X97000529}{\emph{Int.J.Mod.Phys.}
  \textbf{ A12} (1997) 479--534},
  [\href{https://arxiv.org/abs/hep-ph/9512366}{{\texttt hep-ph/9512366}}].

\bibitem{Maniatis:2009re}
M.~Maniatis, \emph{{The Next-to-Minimal Supersymmetric extension of the
  Standard Model reviewed}},
  \href{http://dx.doi.org/10.1142/S0217751X10049827}{\emph{Int. J. Mod. Phys.
  A} \textbf{ 25} (2010) 3505--3602},
  [\href{https://arxiv.org/abs/0906.0777}{{\texttt 0906.0777}}].

\bibitem{Ellwanger:2009dp}
U.~Ellwanger, C.~Hugonie and A.~M. Teixeira, \emph{{The Next-to-Minimal
  Supersymmetric Standard Model}},
  \href{http://dx.doi.org/10.1016/j.physrep.2010.07.001}{\emph{Phys. Rept.}
  \textbf{ 496} (2010) 1--77}, [\href{https://arxiv.org/abs/0910.1785}{{\texttt
  0910.1785}}].

\bibitem{Krawczyk:2002df}
M.~Krawczyk, \emph{{Precision muon g-2 results and light Higgs bosons in the
  2HDM(II)}}, {\emph{Acta Phys. Polon. B} \textbf{ 33} (2002) 2621--2634},
  [\href{https://arxiv.org/abs/hep-ph/0208076}{{\texttt hep-ph/0208076}}].

\bibitem{Domingo:2008bb}
F.~Domingo and U.~Ellwanger, \emph{{Constraints from the Muon g-2 on the
  Parameter Space of the NMSSM}},
  \href{http://dx.doi.org/10.1088/1126-6708/2008/07/079}{\emph{JHEP} \textbf{
  07} (2008) 079}, [\href{https://arxiv.org/abs/0806.0733}{{\texttt
  0806.0733}}].

\bibitem{Mohapatra:1986aw}
R.~Mohapatra, \emph{{Mechanism for Understanding Small Neutrino Mass in
  Superstring Theories}},
  \href{http://dx.doi.org/10.1103/PhysRevLett.56.561}{\emph{Phys. Rev. Lett.}
  \textbf{ 56} (1986) 561--563}.

\bibitem{PhysRevD.34.1642}
R.~N. Mohapatra and J.~W.~F. Valle, \emph{Neutrino mass and baryon-number
  nonconservation in superstring models},
  \href{http://dx.doi.org/10.1103/PhysRevD.34.1642}{\emph{Phys. Rev. D}
  \textbf{ 34} (Sep, 1986) 1642--1645}.

\bibitem{Bernabeu:1987gr}
J.~Bernabeu, A.~Santamaria, J.~Vidal, A.~Mendez and J.~Valle, \emph{{Lepton
  Flavor Nonconservation at High-Energies in a Superstring Inspired Standard
  Model}}, \href{http://dx.doi.org/10.1016/0370-2693(87)91100-2}{\emph{Phys.
  Lett. B} \textbf{ 187} (1987) 303--308}.

\bibitem{Gogoladze:2008wz}
I.~Gogoladze, N.~Okada and Q.~Shafi, \emph{{NMSSM and Seesaw Physics at LHC}},
  \href{http://dx.doi.org/10.1016/j.physletb.2008.12.068}{\emph{Phys. Lett.}
  \textbf{ B672} (2009) 235--239},
  [\href{https://arxiv.org/abs/0809.0703}{{\texttt 0809.0703}}].

\bibitem{Gogoladze:2012jp}
I.~Gogoladze, B.~He and Q.~Shafi, \emph{{Inverse Seesaw in NMSSM and 126 GeV
  Higgs Boson}},
  \href{http://dx.doi.org/10.1016/j.physletb.2012.11.043}{\emph{Phys. Lett.}
  \textbf{ B718} (2013) 1008--1013},
  [\href{https://arxiv.org/abs/1209.5984}{{\texttt 1209.5984}}].

\bibitem{Wang:2013jya}
W.~Wang, J.~M. Yang and L.~L. You, \emph{{Higgs boson mass in NMSSM with
  right-handed neutrino}},
  \href{http://dx.doi.org/10.1007/JHEP07(2013)158}{\emph{JHEP} \textbf{ 07}
  (2013) 158}, [\href{https://arxiv.org/abs/1303.6465}{{\texttt 1303.6465}}].

\bibitem{Dao:2021vqp}
T.~N. Dao, M.~M\"uhlleitner and A.~V. Phan, \emph{{Loop-corrected Higgs Masses
  in the NMSSM with Inverse Seesaw Mechanism}},
  \href{https://arxiv.org/abs/2108.10088}{{\texttt 2108.10088}}.

\bibitem{Cao:2019aam}
J.~Cao, Y.~He, Y.~Pan, Y.~Yue, H.~Zhou and P.~Zhu, \emph{{Impact of leptonic
  unitarity and dark matter direct detection experiments on the NMSSM with
  inverse seesaw mechanism}},
  \href{http://dx.doi.org/10.1007/JHEP12(2020)023}{\emph{JHEP} \textbf{ 12}
  (2020) 023}, [\href{https://arxiv.org/abs/1903.01124}{{\texttt 1903.01124}}].

\bibitem{Cao_2020}
J.~Cao, L.~Meng, Y.~Yue, H.~Zhou and P.~Zhu, \emph{Suppressing the scattering
  of {WIMP} dark matter and nucleons in supersymmetric theories},
  \href{http://dx.doi.org/10.1103/physrevd.101.075003}{\emph{Physical Review D}
  \textbf{ 101} (apr, 2020) }.

\bibitem{Cao:2019evo}
J.~Cao, J.~Lian, L.~Meng, Y.~Yue and P.~Zhu, \emph{{Anomalous muon magnetic
  moment in the inverse seesaw extended next-to-minimal supersymmetric standard
  model}}, \href{http://dx.doi.org/10.1103/PhysRevD.101.095009}{\emph{Phys.
  Rev. D} \textbf{ 101} (2020) 095009},
  [\href{https://arxiv.org/abs/1912.10225}{{\texttt 1912.10225}}].

\bibitem{Cao:2021lmj}
J.~Cao, Y.~He, J.~Lian, D.~Zhang and P.~Zhu, \emph{{Electron and muon anomalous
  magnetic moments in the inverse seesaw extended NMSSM}},
  \href{http://dx.doi.org/10.1103/PhysRevD.104.055009}{\emph{Phys. Rev. D}
  \textbf{ 104} (2021) 055009},
  [\href{https://arxiv.org/abs/2102.11355}{{\texttt 2102.11355}}].

\bibitem{Baglio:2013iia}
J.~Baglio, R.~Gröber, M.~Mühlleitner, D.~Nhung, H.~Rzehak, M.~Spira et~al.,
  \emph{{NMSSMCALC: A Program Package for the Calculation of Loop-Corrected
  Higgs Boson Masses and Decay Widths in the (Complex) NMSSM}},
  \href{http://dx.doi.org/10.1016/j.cpc.2014.08.005}{\emph{Comput. Phys.
  Commun.} \textbf{ 185} (2014) 3372--3391}.

\bibitem{Graf:2012hh}
T.~Graf, R.~Grober, M.~Muhlleitner, H.~Rzehak and K.~Walz, \emph{{Higgs Boson
  Masses in the Complex NMSSM at One-Loop Level}},
  \href{http://dx.doi.org/10.1007/JHEP10(2012)122}{\emph{JHEP} \textbf{ 10}
  (2012) 122}.

\bibitem{Muhlleitner:2014vsa}
M.~Mühlleitner, D.~T. Nhung, H.~Rzehak and K.~Walz, \emph{{Two-loop
  contributions of the order $ \mathcal{O}\left({\alpha}_t{\alpha}_s\right) $
  to the masses of the Higgs bosons in the CP-violating NMSSM}},
  \href{http://dx.doi.org/10.1007/JHEP05(2015)128}{\emph{JHEP} \textbf{ 05}
  (2015) 128}, [\href{https://arxiv.org/abs/1412.0918}{{\texttt 1412.0918}}].

\bibitem{Dao:2019qaz}
T.~Dao, R.~Gröber, M.~Krause, M.~Mühlleitner and H.~Rzehak, \emph{{Two-loop $
  \mathcal{O} $ ( $ {\alpha}_t^2 $ ) corrections to the neutral Higgs boson
  masses in the CP-violating NMSSM}},
  \href{http://dx.doi.org/10.1007/JHEP08(2019)114}{\emph{JHEP} \textbf{ 08}
  (2019) 114}.

\bibitem{Dao:2021khm}
T.~N. Dao, M.~Gabelmann, M.~M\"uhlleitner and H.~Rzehak, \emph{{Two-loop $
  \mathcal{O} $((\ensuremath{\alpha}$_{t}$ + \ensuremath{\alpha}$_{\lambda}$ +
  \ensuremath{\alpha}$_{\kappa}$)$^{2}$) corrections to the Higgs boson masses
  in the CP-violating NMSSM}},
  \href{http://dx.doi.org/10.1007/JHEP09(2021)193}{\emph{JHEP} \textbf{ 09}
  (2021) 193}, [\href{https://arxiv.org/abs/2106.06990}{{\texttt 2106.06990}}].

\bibitem{Casas:2001sr}
J.~A. Casas and A.~Ibarra, \emph{{Oscillating neutrinos and $\mu \to e
  \gamma$}}, \href{http://dx.doi.org/10.1016/S0550-3213(01)00475-8}{\emph{Nucl.
  Phys.} \textbf{ B618} (2001) 171--204},
  [\href{https://arxiv.org/abs/hep-ph/0103065}{{\texttt hep-ph/0103065}}].

\bibitem{Arganda:2014dta}
E.~Arganda, M.~J. Herrero, X.~Marcano and C.~Weiland, \emph{{Imprints of
  massive inverse seesaw model neutrinos in lepton flavor violating Higgs boson
  decays}}, \href{http://dx.doi.org/10.1103/PhysRevD.91.015001}{\emph{Phys.
  Rev. D} \textbf{ 91} (2015) 015001},
  [\href{https://arxiv.org/abs/1405.4300}{{\texttt 1405.4300}}].

\bibitem{Jegerlehner:2009ry}
F.~Jegerlehner and A.~Nyffeler, \emph{{The Muon g-2}},
  \href{http://dx.doi.org/10.1016/j.physrep.2009.04.003}{\emph{Phys. Rept.}
  \textbf{ 477} (2009) 1--110},
  [\href{https://arxiv.org/abs/0902.3360}{{\texttt 0902.3360}}].

\bibitem{Lavoura:2003xp}
L.~Lavoura, \emph{{General formulae for f(1) ---\ensuremath{>} f(2) gamma}},
  \href{http://dx.doi.org/10.1140/epjc/s2003-01212-7}{\emph{Eur. Phys. J. C}
  \textbf{ 29} (2003) 191--195},
  [\href{https://arxiv.org/abs/hep-ph/0302221}{{\texttt hep-ph/0302221}}].

\bibitem{Patel:2015tea}
H.~H. Patel, \emph{{Package-X: A Mathematica package for the analytic
  calculation of one-loop integrals}},
  \href{http://dx.doi.org/10.1016/j.cpc.2015.08.017}{\emph{Comput. Phys.
  Commun.} \textbf{ 197} (2015) 276--290},
  [\href{https://arxiv.org/abs/1503.01469}{{\texttt 1503.01469}}].

\bibitem{Moroi:1995yh}
T.~Moroi, \emph{{The Muon anomalous magnetic dipole moment in the minimal
  supersymmetric standard model}},
  \href{http://dx.doi.org/10.1103/PhysRevD.53.6565}{\emph{Phys. Rev. D}
  \textbf{ 53} (1996) 6565--6575},
  [\href{https://arxiv.org/abs/hep-ph/9512396}{{\texttt hep-ph/9512396}}].

\bibitem{King:2015oxa}
S.~F. King, M.~Muhlleitner, R.~Nevzorov and K.~Walz, \emph{{Exploring the
  CP-violating NMSSM: EDM Constraints and Phenomenology}},
  \href{http://dx.doi.org/10.1016/j.nuclphysb.2015.11.003}{\emph{Nucl. Phys. B}
  \textbf{ 901} (2015) 526--555},
  [\href{https://arxiv.org/abs/1508.03255}{{\texttt 1508.03255}}].

\bibitem{Chen:2001kn}
C.-H. Chen and C.~Q. Geng, \emph{{The Muon anomalous magnetic moment from a
  generic charged Higgs with SUSY}},
  \href{http://dx.doi.org/10.1016/S0370-2693(01)00651-7}{\emph{Phys. Lett. B}
  \textbf{ 511} (2001) 77--84},
  [\href{https://arxiv.org/abs/hep-ph/0104151}{{\texttt hep-ph/0104151}}].

\bibitem{Arhrib:2001xx}
A.~Arhrib and S.~Baek, \emph{{Two loop Barr-Zee type contributions to
  (g-2)(muon) in the MSSM}},
  \href{http://dx.doi.org/10.1103/PhysRevD.65.075002}{\emph{Phys. Rev. D}
  \textbf{ 65} (2002) 075002},
  [\href{https://arxiv.org/abs/hep-ph/0104225}{{\texttt hep-ph/0104225}}].

\bibitem{Heinemeyer:2003dq}
S.~Heinemeyer, D.~Stockinger and G.~Weiglein, \emph{{Two loop SUSY corrections
  to the anomalous magnetic moment of the muon}},
  \href{http://dx.doi.org/10.1016/j.nuclphysb.2004.04.017}{\emph{Nucl. Phys. B}
  \textbf{ 690} (2004) 62--80},
  [\href{https://arxiv.org/abs/hep-ph/0312264}{{\texttt hep-ph/0312264}}].

\bibitem{Heinemeyer:2004yq}
S.~Heinemeyer, D.~Stockinger and G.~Weiglein, \emph{{Electroweak and
  supersymmetric two-loop corrections to (g-2)(mu)}},
  \href{http://dx.doi.org/10.1016/j.nuclphysb.2004.08.014}{\emph{Nucl. Phys. B}
  \textbf{ 699} (2004) 103--123},
  [\href{https://arxiv.org/abs/hep-ph/0405255}{{\texttt hep-ph/0405255}}].

\bibitem{vonWeitershausen:2010zr}
P.~von Weitershausen, M.~Schafer, H.~Stockinger-Kim and D.~Stockinger,
  \emph{{Photonic SUSY Two-Loop Corrections to the Muon Magnetic Moment}},
  \href{http://dx.doi.org/10.1103/PhysRevD.81.093004}{\emph{Phys. Rev. D}
  \textbf{ 81} (2010) 093004}, [\href{https://arxiv.org/abs/1003.5820}{{\texttt
  1003.5820}}].

\bibitem{Fargnoli:2013zda}
H.~G. Fargnoli, C.~Gnendiger, S.~Pa\ss{}ehr, D.~St\"ockinger and
  H.~St\"ockinger-Kim, \emph{{Non-decoupling two-loop corrections to
  $(g-2)_\mu$ from fermion/sfermion loops in the MSSM}},
  \href{http://dx.doi.org/10.1016/j.physletb.2013.09.034}{\emph{Phys. Lett. B}
  \textbf{ 726} (2013) 717--724},
  [\href{https://arxiv.org/abs/1309.0980}{{\texttt 1309.0980}}].

\bibitem{Fargnoli:2013zia}
H.~Fargnoli, C.~Gnendiger, S.~Pa\ss{}ehr, D.~St\"ockinger and
  H.~St\"ockinger-Kim, \emph{{Two-loop corrections to the muon magnetic moment
  from fermion/sfermion loops in the MSSM: detailed results}},
  \href{http://dx.doi.org/10.1007/JHEP02(2014)070}{\emph{JHEP} \textbf{ 02}
  (2014) 070}, [\href{https://arxiv.org/abs/1311.1775}{{\texttt 1311.1775}}].

\bibitem{Cheung:2009fc}
K.~Cheung, O.~C.~W. Kong and J.~S. Lee, \emph{{Electric and anomalous magnetic
  dipole moments of the muon in the MSSM}},
  \href{http://dx.doi.org/10.1088/1126-6708/2009/06/020}{\emph{JHEP} \textbf{
  06} (2009) 020}, [\href{https://arxiv.org/abs/0904.4352}{{\texttt
  0904.4352}}].

\bibitem{Degrassi:1998es}
G.~Degrassi and G.~F. Giudice, \emph{{QED logarithms in the electroweak
  corrections to the muon anomalous magnetic moment}},
  \href{http://dx.doi.org/10.1103/PhysRevD.58.053007}{\emph{Phys. Rev. D}
  \textbf{ 58} (1998) 053007},
  [\href{https://arxiv.org/abs/hep-ph/9803384}{{\texttt hep-ph/9803384}}].

\bibitem{Barr:1990vd}
S.~M. Barr and A.~Zee, \emph{{Electric Dipole Moment of the Electron and of the
  Neutron}}, \href{http://dx.doi.org/10.1103/PhysRevLett.65.21}{\emph{Phys.
  Rev. Lett.} \textbf{ 65} (1990) 21--24}.

\bibitem{Abe:2013qla}
T.~Abe, J.~Hisano, T.~Kitahara and K.~Tobioka, \emph{{Gauge invariant Barr-Zee
  type contributions to fermionic EDMs in the two-Higgs doublet models}},
  \href{http://dx.doi.org/10.1007/JHEP01(2014)106}{\emph{JHEP} \textbf{ 01}
  (2014) 106}, [\href{https://arxiv.org/abs/1311.4704}{{\texttt 1311.4704}}].

\bibitem{Bechtle:2020pkv}
P.~Bechtle, D.~Dercks, S.~Heinemeyer, T.~Klingl, T.~Stefaniak, G.~Weiglein
  et~al., \emph{{HiggsBounds-5: Testing Higgs Sectors in the LHC 13 TeV Era}},
  \href{http://dx.doi.org/10.1140/epjc/s10052-020-08557-9}{\emph{Eur. Phys. J.
  C} \textbf{ 80} (2020) 1211},
  [\href{https://arxiv.org/abs/2006.06007}{{\texttt 2006.06007}}].

\bibitem{Bechtle:2020uwn}
P.~Bechtle, S.~Heinemeyer, T.~Klingl, T.~Stefaniak, G.~Weiglein and
  J.~Wittbrodt, \emph{{HiggsSignals-2: Probing new physics with precision Higgs
  measurements in the LHC 13 TeV era}},
  \href{http://dx.doi.org/10.1140/epjc/s10052-021-08942-y}{\emph{Eur. Phys. J.
  C} \textbf{ 81} (2021) 145},
  [\href{https://arxiv.org/abs/2012.09197}{{\texttt 2012.09197}}].

\bibitem{Zyla:2020zbs}
{\scshape Particle Data Group} collaboration, P.~Zyla et~al., \emph{{Review of
  Particle Physics}}, \href{http://dx.doi.org/10.1093/ptep/ptaa104}{\emph{PTEP}
  \textbf{ 2020} (2020) 083C01}.

\bibitem{Aoyama:2017uqe}
T.~Aoyama, T.~Kinoshita and M.~Nio, \emph{{Revised and Improved Value of the
  QED Tenth-Order Electron Anomalous Magnetic Moment}},
  \href{http://dx.doi.org/10.1103/PhysRevD.97.036001}{\emph{Phys. Rev. D}
  \textbf{ 97} (2018) 036001},
  [\href{https://arxiv.org/abs/1712.06060}{{\texttt 1712.06060}}].

\bibitem{Hanneke_2008}
D.~Hanneke, S.~Fogwell and G.~Gabrielse, \emph{New measurement of the electron
  magnetic moment and the fine structure constant},
  \href{http://dx.doi.org/10.1103/physrevlett.100.120801}{\emph{Physical Review
  Letters} \textbf{ 100} (mar, 2008) }.

\bibitem{PhysRevA.83.052122}
D.~Hanneke, S.~Fogwell~Hoogerheide and G.~Gabrielse, \emph{Cavity control of a
  single-electron quantum cyclotron: Measuring the electron magnetic moment},
  \href{http://dx.doi.org/10.1103/PhysRevA.83.052122}{\emph{Phys. Rev. A}
  \textbf{ 83} (May, 2011) 052122}.

\end{thebibliography}
\end{document}